\documentclass[12pt,preprint]{aastex}

\newcommand\lam{\mbox{$\:\lambda $ }}
\newcommand\lamlam{\mbox{$\:\lambda\lambda $ }}
\newcommand\ha{{H$\alpha$}}

\newcommand\kms{\:\rm{\,km\,s^{-1}}}

\newcommand\perpix{\:{\rm pixel}^{-1}}
\newcommand\LUM{\:{\rm ergs\:s^{-1}}}
\newcommand\FLUX{\:{\rm ergs\:cm^{-2}\:s^{-1}}}
\newcommand\FLUXARCSEC{\:{\rm ergs\:cm^{-2}\:s^{-1}\:arcsec^{-2}}}

\newcommand\hi{\ion{H}{1}}

\newcommand\sii{[\ion{S}{2}]}
\newcommand\oi{[\ion{O}{1}]}
\newcommand\oii{[\ion{O}{2}]}
\newcommand\oiii{[\ion{O}{3}]}
\newcommand\nii{[\ion{N}{2}]}

\newcommand\tabsii{[S\thinspace II]}
\newcommand\taboi{[O\thinspace I]}
\newcommand\taboii{[O\thinspace II]}
\newcommand\taboiii{[O\thinspace III]}
\newcommand\tabneiii{[Ne\thinspace III]}

\slugcomment{To appear in {\it The Astronomical Journal}, Vol. 132, July 2006}

\shorttitle{Ejecta Filaments in SNR G292.0+1.8 }
\shortauthors{Winkler \& Long}

\begin{document}


\title{Far-flung Filaments of Ejecta in the Young Supernova Remnant
G292.0+1.8}


\author{P. Frank Winkler\altaffilmark{1}}  
\affil{Department of Physics, Middlebury College, Middlebury, VT 05753}
\email{winkler@middlebury.edu}
\and
\author{Knox S. Long\altaffilmark{1}}
\affil{Space Telescope Science Institute, 3700 San Martin Drive, Baltimore MD 21218}
\email{long@stsci.edu}

\altaffiltext{1}{Visiting Astronomer, Cerro Tololo Inter-American Observatory.
CTIO is operated by AURA, Inc.\ under contract to the National Science
Foundation.}


\begin{abstract}
New optical images of the young supernova remnant (SNR) G292.0+1.8, obtained 
from the 0.9-m telescope at CTIO, show a more extensive network  
of filaments than had been known previously.   Filaments emitting in \oiii\  are distributed
throughout much of the 8 arcmin diameter shell seen in X-ray and radio images, including
a few at the very outermost shell limits.  
In addition to the extensive \oiii\ filaments, we have detected four small complexes
of filaments that show \sii\  emission along with the oxygen lines.   
In a single long-slit spectrum we find variations of almost an order of magnitude in the
relative strengths of oxygen and sulfur lines, which must result from abundance 
variations.  None of the filaments, 
with or without \sii, shows any evidence for hydrogen, so all appear to  be  
fragments of pure supernova ejecta.   The \sii\  filaments provide the first evidence 
for undiluted products of oxygen burning in the ejecta from the supernova that gave rise to 
G292.0+1.8.    Some oxygen burning, either hydrostatic or explosive, must have occurred, 
but the paucity of \sii -emitting filaments suggests that either the oxygen burning was not
extensive or that most of its products have yet to be excited.
Most of the outer filaments exhibit radial, pencil-like morphologies that suggest 
an origin as Rayleigh-Taylor fingers of ejecta, perhaps formed during the explosion.   
Simulations of core-collapse 
supernovae predict the development of such fingers, but these have never before been 
so clearly observed in a young SNR. 
Following careful subtraction of the stars in the field, we have measured the total
flux in \oiii \lam 5007 as $5.4 \times 10^{-12} \FLUX$.   Using a distance of 
6 kpc and an extinction 
correction corresponding to $E(B-V) = 0.6$\ mag (lower than previous values but more consistent
both with our data and with X-ray and radio measurements of the hydrogen column 
density)  leads to a luminosity of $1.6 \times 10^{35} \LUM $\ 
in the 5007 \AA\ line.

\end{abstract}




\section{Introduction}

The canonical picture of a supernova remnant (SNR) that results from the core collapse 
and supernova explosion of a
massive star includes a shell produced by an expanding shock, fragments of ejecta
that may appear as optical filaments rich in oxygen and (often) other heavy elements, a central
neutron star that manifests itself as a pulsar, and a synchrotron-emitting pulsar wind nebula 
that surrounds and is powered by the pulsar.   Yet there is only a single
SNR in the Galaxy that has been found to display all these 
properties:  G292.0+1.8, also known as MSH 11--5{\it 4}, 
and its associated pulsar PSR J1124--5916 \citep{hughes01, camilo02, hughes03}.   
This system thus provides a unique opportunity
to study both the compact and ejected remains from a massive star and their interaction with the
local environment.

Although G292.0+1.8 (henceforth more succinctly G292) was originally detected 
and identified as a SNR through its radio properties  \citep{mills61, milne69}, it
first attracted significant attention when  \citet{goss79} discovered optical filaments with
spectra dominated by lines of oxygen and neon.   Hydrogen Balmer lines, and also 
lines of \sii\ and \nii\  typically seen in SNRs, were conspicuous by
their absence.   \citet{murdin79} showed these filaments to have high radial 
velocities, $-700 \kms \lesssim v_{rad} \lesssim +1300 \kms$, indicating that they 
are undecelerated (or minimally decelerated) ejecta from the supernova explosion.
G292 thus joined the prototype Cas A in the 
class of ``oxygen-rich'' SNRs---an exclusive group that still comprises 
only about 8 members.  The optical filaments with extremely metal-rich spectra 
that characterize the O-rich remnants 
are fragments of nearly pure ejecta that were launched from the core of the progenitor star during 
its explosion and  that remain virtually uncontaminated through interaction with 
interstellar or circumstellar material.    Subsequent spectra of G292 by \citet{dopita84} and 
\citet{sutherland95a} have shown no lines other than those of oxygen and neon.

G292 has been the subject of numerous X-ray studies, culminating in the spectacular 
high-resolution image from {\it Chandra} \citep{park02}.   
The X-ray spectrum is dominated by K-shell lines of O, Ne, Mg, Si, and S.  Non-equilibrium 
ionization analyses require large enhancements (relative to solar) in abundances 
for O, Ne, and Mg, with lesser enhancements for Si, S, and Fe \citep{hughes94, gonzalez03}.
By comparing the inferred abundances with the integrated yields predicted by  models 
for core-collapse supernovae, they estimated a progenitor mass of $\sim20-40\:M_\sun$.
In a more detailed study of individual X-ray features,  \citet{park04} found that
different knots have very different  compositions and suggested that  these 
represent clumps of ejecta from different zones of the progenitor.  

The {\it Chandra} image of G292 also revealed a compact central source, 
surrounded by what appeared to be a pulsar wind nebula \citep{hughes01}.  
Shortly thereafter, \citet{camilo02} discovered the radio pulsar PSR J1124--5916 within
or very near G292.   \citet{hughes03} then showed that the compact X-ray source is pulsed with the same period, confirming that it must be the compact remnant of the star that produced G292.  
The period of 135 ms and 
spin-down age of 2900 yr are roughly consistent with the age of the G292 SNR, estimated
by \citet{chevalier05} as 2700--3700 yr based on properties of the pulsar-wind nebula, 
and more directly as 3000--3400 yr from the kinematic study by \citet{ghavamian05}.

Detailed radio images of G292  obtained by \citet{gaensler03} show a bright central 
core around the pulsar, surrounded by a fainter plateau about 8\arcmin\ in diameter with sharp
outer edges.   The outer edge of the radio plateau coincides closely with the outer extent
of X-ray emission seen in the Chandra image, and must delineate the SNR's primary 
shock.  Based primarily on the \hi\ absorption profile, 
\citet{gaensler03} also give what is probably the most reliable distance estimate
to G292:  $6.2\pm 0.9\;$ kpc, which implies a diameter of about 15 pc for the SNR.

In this paper we present the first CCD images that cover the full extent of G292.  
The narrow-band  \oiii\ image, especially after we subtract  
a matched continuum image, shows a network of filaments 
far more extensive than indicated in photographic studies \citep{goss79, tuohy82}, and 
even more wide-spread
than those in the Fabry-Perot studies of \citet{ghavamian05}.
In addition, we have discovered a few filaments that show emission lines of \sii\ in addition to oxygen and neon---the first evidence for the presence of oxygen-burning products in the ejecta-dominated optical filaments.   

\section{Emission-Line Images}
We have obtained optical images of G292 in emission lines of \oiii, \sii, and \ha, along with matched red and green continuum bands, from CTIO using the 0.9-m telescope and Tek2K no. 5 CCD  
on 2002 March 20 and 23 (UT).  
The continuum bands are important since they enable us to subtract away most of the light from the
numerous stars that litter this dense Galactic field to reveal faint, small scale emission features of 
the SNR itself.  The  $2048\times2048$\ pixel chip covered a field 13\farcm8 square at a scale of $0\farcs401 \perpix$, easily accommodating the $\sim 8\arcmin$ diameter of G292.  We were favored with photometric conditions and seeing about 1\arcsec\ throughout the run, and all the G292 images were taken with no moon.  We obtained 3 to 5 frames in each filter and dithered the telescope 
by a few arcsec between frames to paper over cosmetic defects and to minimize systematic
effects of chip sensitivity.

Details of the observations are given in Table 1.   The images were processed using standard 
IRAF\footnote{IRAF is distributed by the National Optical Astronomy 
Observatories, which is operated by the  AURA, Inc. under cooperative 
agreement with the National Science Foundation.}  procedures of bias-subtraction and flat-fielding.   Flux calibration was achieved through observation of spectrophotometric standard stars from the list of \citet{hamuy92} each night during the run.   All the images were aligned to a common coordinate system using some 300 stars from the UCAC1 catalog \citep{ucac1_00}. 

The combined \oiii\  image is shown in Fig.\ 1, both before (Fig.\ 1a) and after (Fig.\ 1b) 
continuum subtraction.  
Although artifacts of many bright stars remain, the faint emission features are far more evident
in the continuum-subtracted image.   In Fig.\ 2 we compare  our \oiii\ image of G292 
with an X-ray image from the {\it Chandra} ACIS-S image on the same scale.   (These data, 
obtained from the {\it Chandra} archive, are the same as those used by \citet{park02} for their 
color image.)  We have placed matching contours on both images to clarify the relative 
locations of X-ray and optical features.    

\notetoeditor{We request that the first six figures, 1a, 1b, 2a, 2b, 3, and 4, be reproduced at the
same scale.  We think that this should probably be wider than single-column.  It would be 
particularly useful to have Figs 1a. and 1b. appear either one beneath the other or on facing 
pages, directly opposite one another.  Similarly for Figs 2a and 2b.  (We have no objection
to placing all four of these figures on a pair of facing pages, but this is not essential.)
And finally it would be useful but not essential to have Figs 3 and 4 on the same or facing pages
where they can be easily compared with one another.}

It is apparent  from these CCD images  that the \oiii -emitting filaments in G292 
are far more numerous and more extensive than early photographic images  \citep{goss79, tuohy82}\footnote{When comparing
our images with those of \citet{goss79}, note that their Plate 1 is reproduced as a mirror image
to the conventional north up, east left orientation.} indicated.  Not surprisingly, features in the central 
region coincide with those shown in the Fabry-Perot images of \citet{ghavamian05}, but 
our images also show many filaments near the periphery of the shell, outside 
their Fabry-Perot field.  In addition to the 
bright crescent-shaped group of filaments east of the X-ray center, sometimes referred to as the ``spur,"
fainter individual knots 
and more diffuse filaments are present throughout much of the  $8\arcmin$\  diameter of 
the X-ray remnant.   

It might at first appear that the southernmost optical filaments, located 
near $\alpha(2000) = 11^{\rm h}24^{\rm m}29^{\rm s},\ 
\delta (2000) = -59\degr 20\arcmin 38\arcsec$, lie outside the X-ray shell.  
This is not the case; these filaments simply lie outside the field covered by the ACIS-S3 
chip.   X-ray images from the {\it ROSAT} HRI  (unpublished archive data), which cover a wider 
field than the ACIS image (though with lower sensitivity and angular resolution), show 
a diffuse X-ray extension at the southern extreme of the G292 shell.  We have added the
outer contours from the {\it ROSAT} image to Fig.\  2b in order to show the full extent of X-ray
structure.  Still, the 
southernmost \oiii\  filaments appear to lie very near the outer extreme of even this 
extended X-ray emission.  


In addition to the extensive network of filaments seen in the \oiii\ images, we have also
detected a few filaments in \sii\ $\lambda\lambda$\ 6716, 6731.  
Our \sii\ images show at least four faint individual filaments or small complexes, illustrated in Fig.\ 3.  
These represent the first evidence for the presence of sulfur, or of any of the Si-group elements that result from O-burning, in optical filaments of  G292\@.   Though clear and unambiguous, the \sii\ emission is significantly fainter than that in \oiii\@.  The brightest \oiii\ filament is the northern portion of  
the spur, which runs  nearly E-W
centered at  $\alpha(2000) = 11^{\rm h}24^{\rm m}48\fs 7,\ 
\delta (2000) = -59\degr 15\arcmin 41\arcsec$\@.  In the very brightest area 
of this filament 
the \oiii\ intensity is $2.8 \times 10^{-15}\FLUXARCSEC$, more than  
20 times the \sii\ intensity at the same location.  
An extension of Filament 1,  located $\sim 18\arcsec$\  to the west and 
designated 1W in Fig.\ 3, has $I_{\rm [S\; II]} = 0.5  \times 10^{-15}\FLUXARCSEC$, while  
$I_{\rm [O\; III]} = 1.6 \times 10^{-15}\FLUXARCSEC$\  is now only 3 times brighter.   
The slit for the spectrum presented in \S\thinspace 3 was oriented to cover both the above 
portions of Filament 1, designated 1 East and 1 West, as shown in detail in Fig.\ 5.  

The brightest of the \sii\ filaments is the one designated 3 in Fig.\ 3 , and located at 
$\alpha(2000) = 11^{\rm h}24^{\rm m}39\fs 8,\ 
\delta (2000) = -59\degr 12\arcmin 36\arcsec$.
Marginally resolved in our images, it has $I_{\rm [S\; II]} \approx 1.0 \times 10^{-15}\FLUXARCSEC$, more than twice $I_{\rm [O\; III]} $\ at the same location.  There are no spectra of this filament
as yet.  

We can estimate the integrated \oiii\ flux from the entire remnant by summing the flux in each of the regions containing filaments, to obtain $F_{5007} \approx 5.4 \times 10^{-12} \FLUX$, with an
uncertainty which we estimate as $\sim 20\%$\@. The myriad stars in the field
contribute many times more (continuum) flux in the 60-\AA\  bandpass than the line flux from the filaments, so imperfect star subtraction can lead to a significant error in the nebular flux.   With this caveat we give our result, since we know of no previous values for such a quantity from G292.

Finally, we have also obtained an image in \ha,  shown in Fig.\ 4.   All  of the 
fine filaments seen in \oiii\  or \sii\  are completely absent in \ha.  There is a considerable amount
of faint diffuse \ha\ structure, virtually all of which also appears even more faintly in \sii\ (Fig.\ 3).
Figure 9 of \citet{ghavamian05} shows the \ha\ emission in the central region of G292 
more clearly, and these authors further show that all the \ha\ is 
at essentially zero radial velocity, quite unlike the \oiii\ filaments.  They 
speculate that the \ha\ structures may result from interstellar or circumstellar material in 
the vicinity of G292 that has been excited either by slow shocks or by photoionization 
caused by the strong radiative shocks that produce the X-ray emission.   
Looking at these structures in a larger context, however, shows that  diffuse
structures  seen in both \ha\ and \sii, very similar to ones within G292, extend well
beyond the boundary of the SNR shell.  It is entirely possible that this  diffuse emission may not
be related to G292 at all.
Not surprisingly, there is considerable diffuse \ha\ emission on a much larger scale throughout
this field near the Galactic plane.  The SHASSA survey \citep{gaustad01} shows a band of diffuse 
emission passing roughly 
NE to SW through G292, with $I_{\rm H\alpha} \sim 80$\ to 90 Rayleigh 
(averaged over the $0\farcm 8 \perpix$\
resolution of that survey), equivalent to an emission 
measure of 180--200 cm$^{-6}$ pc.   
The intensity of the brightest diffuse \ha\ within the G292 shell is 
$I_{\rm H\alpha} = 0.6  \times 10^{-15}\FLUXARCSEC = 110\;$ Rayleigh, which represents   
little if any enhancement over the diffuse Galactic emission seen throughout the region.
The question of whether or not the \ha\ emission is physically associated with G292 may 
eventually be resolved through observations from {\it Spitzer}, since one would
expect mid-infrared radiation from dust grains heated by the passage of a strong SNR shock.

Since the ejecta-dominated filaments are known to have high velocities, the pass bands of 
the filters are important, and the velocity ranges are included in Table 1.  The \oiii\ filter 
admits emission in the velocity range --1700 to +1800$\kms $, easily encompassing the 
full range for which emission has been reported for G292 by \citet{ghavamian05} or any
previous studies.   The \sii\  image should include 
emission in one or both lines of the 
$\lambda\lambda $\ 6716, 6731 doublet from material in the range --1200 to +1600$\kms$,  
but it is possible that extremely blue-shifted \sii\ emission might have fallen largely outside 
the pass band of our filter.   The same is true for \ha, of course, where the pass band is 
narrower ($-450\ {\rm to}\, + 650 \kms$),  
but neither we nor \citet{ghavamian05} see any evidence for \ha\ emission from  any  of the fast, ejecta-dominated filaments.

\section{Long-slit Spectroscopy}
We obtained deep spectra of filament 1, the brightest filament in G292, 
from the CTIO 1.5m telescope and R-C spectrograph in 1997 March.\footnote{This 
was 5 yr prior to the run during which the image data presented in \S\thinspace 2
were obtained.  We had found evidence for [S\thinspace II] emission during an earlier imaging 
observation that covered only the central region of G292 and had poorer seeing than 
the images shown in \S\thinspace 2.}
A slit of width 3\farcs5 ($193.5\;\mu$) was aligned 
at a position angle of 100\fdg5 as shown in Fig.\ 5,
in order to capture the brightest portion of the \oiii\ and the 
\sii\ emission in spatially distinct regions along the slit.
Three different spectrograph setups were used, as detailed in Table 2.  All used 300 line/mm gratings to give a dispersion of $2.9\: {\rm\AA} \perpix$\@.  The observations using the red and IR setups were 
carried out on the night of 1997 March 28 (U.T.), and those in the blue setup were done the 
following night.   We were careful to adjust the slit position and orientation to be virtually identical  for 
all  three spectra.

Data reduction was carried out using IRAF and following standard CTIO procedures for this spectrograph:
initial flat-fielding using dome flats (combined with an internal quartz-lamp flat for the 
blue setup only) was followed by slit-illumination correction using well-exposed twilight 
sky flats.  Comparison spectra using a HeNeAr lamp were obtained both before
and after each sequence of observations for wavelength calibration.   For flux calibration
we used several stars from the \citet{hamuy92} list, observed over a range of airmasses.

The two-dimensional spectrum from the red setup is shown in Fig.\ 6.  Emission from Filament 1
shows a similar characteristic curvature in each of several emission lines, the result of a radial 
velocity gradient along the filament.   The complete absence of any \ha\ emission showing
a similar velocity profile constitutes further evidence that this filament is composed almost entirely  
of heavy elements, i.e., supernova ejecta.
Close inspection of the figure shows that the 2-dimensional
emission pattern is virtually identical in lines of \oi, \oii, and \oiii, but that the \sii\ lines show
a somewhat different pattern.  All of the oxygen lines are strongest toward the eastern end
of the filament (toward the bottom in Fig.\ 6),
while the \sii\ lines are strongest in the west .  This
is exactly what one expects from comparing the images in \oiii\ and \sii\ (Fig.\ 5).  

Fig.\ 7 shows a pair of one-dimensional spectra obtained by summing over several rows in two
regions along the slit:  one to the east where the oxygen lines are strongest 
(denoted 1E in Fig.\ 6),
and the other to the west  where the \sii\ lines
are strongest (denoted 1W).
These one-dimensional data shown in Fig.\ 7 are actually the combination of spectra
extracted from identical regions of all three spectrograph setups.  (We do not show the spectrum at 
$\lambda > 8000\: {\rm \AA}$, since no lines emerge above the noise at these wavelengths.) 
Furthermore, we applied a transformation to
artificially ``straighten'' the spectra before extraction, in order to avoid broadening the lines 
by summing emission from material at different velocities.    The line fluxes in the spectra
of the two regions are tabulated in Table 3.

There has been no optical spectroscopy of G292 capable of yielding an unambiguous determination
of the reddening, but virtually all of the scant literature describing G292's optical emission has 
adopted the value of  $E(B-V) = 0.9$\ mag proposed by \citet{goss79}.  We argue in
\S\thinspace 4.2 that spectra both in the optical and in other bands are best interpreted by using 
a smaller value,  $E(B-V) \approx 0.6$\ mag.  In Table 3 we give dereddened intensities
for both these values of $E(B-V)$, where we have used the extinction function of
\citet{cardelli89} with $A(V)/E(B-V) = 3.1$.

Although oxygen lines clearly dominate the spectrum from both regions, the \sii\ lines are 
stronger {\it relative to the oxygen lines} by about a factor of 8 in the western (upper) spectrum  
when compared with the eastern (lower) one. 
The relative strengths of the \oi, \oii, and \oiii\ lines are similar in the two
spectra, therefore we can exclude the possibility that the variation in S/O line strengths 
along the filament results from some difference the in physical  conditions.  Instead, variations in the S/O 
abundance ratio must be responsible.  
Furthermore, note that in the east the \sii\ lines suggest an electron density near the 
low-density limit, while the much stronger \sii\ lines in the west indicate $n_e \sim 200\; {\rm cm}^{-3}$.

Finally, we note that in addition to the numerous forbidden lines of oxygen, 
{\it permitted} \ion{O}{1} \lam 7774  is 
also present.  This is a recombination line, and 
the fact that it is not just present, but stronger than \ha, indicates that
O$^+$\  is winning out over H$^+$\ in the competition for recombination electrons; 
i.e., that oxygen is more abundant
than hydrogen.  The same line has been observed in other oxygen-rich SNRs such as 
Puppis A \citep{winkler85}, 
Cas A \citep{winkler91, hurford96}, and the SMC remnant E0102.2--7219 \citep{blair89}.

\section{Discussion}

\subsection{Reddening, Spectral Diagnostics, and \oiii\  Luminosity}

\citet{goss79}, in the paper where  G292 was first identified as an oxygen-rich
SNR, estimated that the reddening was 
$E(B-V) = 0.9$\ mag, arguing for consistency in the relative strengths of the oxygen lines.
Although this argument depends entirely on highly uncertain models  
for the emission from shocked material 
composed almost entirely of metals, this original value for the reddening has persisted in the literature.  
An alternative approach to obtaining the reddening is through the hydrogen column density, $N_H$,
which \citet{predehl95} have shown to be closely correlated with optical extinction.
They found that these quantities to be related by 
$N_H/A_V = 1.79 \times 10^{21} \;{\rm cm^{-2}mag^{-1}}$, equivalent to 
$E(B-V) = 5.55 \times 10^{21} \;{\rm cm^{-2}mag^{-1}}$.

There have been a number of recent measurements of $N_H$\ based on X-ray and
radio data.  \citet{gonzalez03} found a value $(5 \pm 1) \times 10^{21}\; {\rm cm^{-2}}$\ 
based on an average of fits to the spectra from numerous X-ray knots in G292, suggesting
$E(B-V) = 0.9 \pm 0.2$\ mag.  \citet{park04} fit the spectra of a number of individual
X-ray knots and found values for $N_H$\ ranging from 3 to 7.7 $\times 10^{21}\; {\rm cm^{-2}}$,
equivalent to $E(B-V)$\ from 0.5 to 1.4 mag.  However, modeling of thermal X-ray 
plasmas with abundances that may be far from solar is difficult, and the absolute accuracy
of fits based on these models is hard to assess.  For this reason, we believe the most
reliable X-ray determination of the column density is that by \citet{hughes01},
who fit  the pulsar spectrum with an absorbed power law to obtain
$N_H \approx (3.17 \pm 0.15) \times 10^{21}\; {\rm cm^{-2}}$, close to the minimum
value derived by \citet{park04}.  This lower value agrees well with the column obtained 
using radio data by \citet{gaensler03}, who  
integrated the \hi\ emission over a range of velocities and inferred
that $N_H \approx 3.3 \times 10^{21}\; {\rm cm^{-2}}$, equivalent to $E(B-V)=0.6$\ mag.

The most successful models for the optical emission from young, oxygen-rich SNRs are
those of \citet{sutherland95b}, who calculated the emission resulting when a 
strong (reverse) shock enters an oxygen-rich fragment of material from a progenitor core.  
They considered both direct post-shock emission 
and also the indirect effects of  X-rays that can  produce an
ionization front propagating into the fragment ahead of the shock itself.
They used a grid of models considering either or both of these processes, 
with a range of shock velocities, and calculated 
diagnostic ratios for various oxygen lines.
Comparing  their models with our observed spectra,  we find that 
dereddening as much as $E(B-V) = 0.9$\ leads to intensities
for  \oii \lam3727 that are stronger than any of their models predict.   
\citet{sutherland95b} had precisely the same problem for G292, and indeed their
own observed intensity ratios \citep[from][where they also used  $E(B-V) = 0.9$]{sutherland95a}
are almost exactly the same as ours for the bright Filament 1 East.   
But decreasing the reddening to $E(B-V) = 0.6$\ gives ratios very similar to 
their model OSP150, which includes both direct post-shock emission and 
preionization from a $150 \kms$\ shock, and also similar to ratios they report
for a number of filaments in the O-rich remnant N132D in the Large Magellanic Cloud.  
For example, the most reddening-sensitive ratio in their analysis is 
\oii \lam7325:\oii \lam3727.  For $E(B-V) = 0.6$, our dereddened intensity ratio
for both the East and West regions of Filament 1 (see Table 3) is
0.028, almost exactly that predicted by the OSP150 model, while for
$E(B-V) = 0.9$\ it is half this value and substantially lower than has been observed
in dereddened spectra of other O-rich SNRs \citep[see e.g.][Fig.\ 13c]{sutherland95b}.

Here and in the subsequent discussion we adopt the value $E(B-V)=0.6$\ mag, 
which is consistent with most X-ray and radio measurements of the column density
and gives relative line strengths that are consistent with the best models for a 
plasma composed entirely of heavy elements.  This in turn allows us to estimate the
luminosity of G292 in \oiii \lam 5007.  Using our estimate (\S\thinspace 2) of the integrated flux,
$F_{5007} = 5.4 \times 10^{-12} \FLUX$, 
and our adopted value for the extinction implies 
$L_{5007} = 1.6 \times 10^{35}\:d_6^{\:2}\; \LUM$, where $d_6$\ is the distance modulo 6 kpc.  
This luminosity value has an uncertainty $\sim 20\%$\ 
attributable to our flux estimate, in addition to any uncertainty in the extinction.

One other O-rich SNR for which the \oiii\ luminosity has been reported is 
the extremely bright SNR in the Magellanic-type irregular galaxy NGC 4449, at 
a distance of about 3.9 Mpc \citep{hunter99}.  For this case \citet{blair83}\footnote
{The 4959 \AA\ and 5007 \AA\ lines in the NGC 4449 SNR are so broad that 
they cannot be resolved.  Note also that \citet{blair83} used a distance of 5 Mpc and thus 
found even higher values for the luminosity.}
give $F_{4959+5007} = 2.02 \times 10^{-13} \FLUX$\@.  If we attribute $3/4$\ of the total flux to the 5007 \AA\ line and correct after for 
extinction we find  $L_{5007} = 5.2 \times 10^{38} \LUM$, 300 times higher than G292.
It has been recognized since its discovery, however, that the NGC 4449 SNR is extraordinary; 
its X-ray and radio luminosities are also higher than any other known remnants.  
It should be straightforward to obtain \oiii\ luminosities for most or all of the other O-rich
SNRs, but no values are readily available in the literature. 

\subsection{\sii\ Emission and Nucleosynthesis Implications}

The detection of \sii\ emission, even if in only a handful of filaments, is nevertheless
significant.   Sulfur resulting from massive-star
supernovae stems almost entirely from
oxygen burning, either hydrostatically in a shell surrounding the core during a
short time before core collapse, and/or explosively during the supernova itself 
\citep[e.g.][]{arnett96,woosley02}.   
The presence of S in fragments of pure SN ejecta (which these filaments must be since
they are devoid of H)
represents definitive evidence that some oxygen burning has occurred in the
G292 progenitor.  The apparent absence of S or other O-burning products
in optical spectra from G292 had been something of a puzzle, since 
models for the yields from 
core-collapse SNe all  predict from 0.04 to 0.2 $M_\sun$\ of S \citep[e.g.][]{rauscher02}. 

An analysis of the abundances in  filaments composed entirely of heavy
elements represents a challenge beyond the scope of this paper;  
nevertheless, the detection of only relatively faint \sii\ emission in a small minority of G292's 
filaments clearly indicates that S 
is not a prominent constituent in the optical filaments of this SNR\@.    
Either there has been 
only a little oxygen burning, or most of the O-burning products have yet to be excited.
Comparing G292 with other O-rich SNRs, only two of these 
have strong \sii\ lines in their optical spectra:  
Cas A \citep[e.g.][]{chevalier78, fesen01} and 0540--69.3 in the Large Magellanic Cloud
\citep{dopita84, kirshner89}.   In the other members of this small society---
Puppis A \citep{winkler85,sutherland95a}, N132D in the LMC and 1E0102--7219 in the SMC
 \citep[e.g.][for both]{blair00}, the very bright SNR in NGC 4449 \citep{blair83}, and the remnant of 
 SN1957D in M83 \citep{long89}--- \sii\ is 
extremely weak or absent entirely in the ejecta-dominated filaments.   

\subsection{Optical--X-Ray Comparison}

Comparison of the optical images with high-resolution X-ray images from {\it Chandra} 
\citep{park02} shows
little detailed correspondence in features.  The brightest optical feature, filament 1 where
we obtained our spectrum, has no exact X-ray counterpart; it lies $\sim 30\arcsec $\ north of the 
eastern end of the bright equatorial X-ray belt.   This X-ray belt does not show enhanced 
abundances, and \citet{park02} have interpreted it as a region where the primary supernova
shock is encountering a ring of circumstellar material, seen approximately edge-on.  
The proximity of such a bright, ejecta-dominated  filament to one end of this projected  
ring may result from a strong reverse shock, produced indirectly by the encounter with
the circumstellar ring, propagating  into ejecta expanding roughly transverse to 
our line of sight.  The low radial velocity observed for this filament, $v_r \approx 100 \kms$, 
is consistent with near-transverse motion.    

\citet{park02} show equivalent-width (EW) maps for strong X-ray lines of several metals.  
For oxygen, the highest equivalent widths are found in a complex within the 
southeast quadrant of G292.    The largest number of O-rich optical filaments are located
along a broad arc extending  through this quadrant roughly from NNE to SSW (Fig.\ 1).  
The Ne (He $\alpha$) and Mg X-ray EW maps also show strong emission along
a similar arc.  This suggests that pure ejecta fragments similar in composition to the
ones we now observe have provided the material for the X-ray plasma in this region.

The {\it Chandra} data partially complement the optical picture of  
oxygen-burning products inhomogeneously distributed around G292.   The \citet{park02}
EW map for Si shows a string of prominent X-ray knots extending almost two-thirds of
the way across the northern portion of the shell.   The two optical filaments where the 
S/O ratio is highest are both found in this general region.  Filament 4 lies near the center 
of a relatively bright X-ray feature, designated as Region 5 by \citet{park04} who found
this region to be relatively more enhanced in Si and S than other portions of the SNR\@.
Our filament 3 lies about 40\arcsec\ to the NW, near the edge of another X-ray complex. 
Fig.\ 8 shows a detailed optical/X-ray comparison for this portion of G292. 
On the version of the {\it Chandra} image by \citet{gonzalez03}, filaments 4 and 3 are 
located within their regions 16 and 17, respectively.   These contribute to the global
abundances they calculated, but they did not report line fluxes or abundance 
estimates for individual regions.

A particularly interesting set of  O-rich filaments is at the extreme south, 
outside the field of the {\it Chandra} ACIS image and at the extreme edge of
X-ray emission detected from the {\it ROSAT} HRI, as pointed out in \S\thinspace 2.
Fragments of ejecta 
launched in this direction have traveled farther than any others, so these are 
probably the fastest such fragments in G292.   
It appears that  the primary shock is encountering a low-density region 
to the south, leading to more rapid expansion in this direction.  Perhaps pre-supernova
mass loss has swept out the ISM in this direction, perpendicular to the equatorial
ring seen in X-rays.   One might suspect a bipolar flow, leading to similar phenomena to
the north, but as yet there is no evidence for any emission beyond the near-circular
northern portion of the shell.   It would be interesting to obtain optical spectra of 
one or more of the extreme southern filaments and to investigate the X-ray structure here in more 
detail than can is possible with the {\it ROSAT} data.

It is interesting to compare the total X-ray luminosity of G292 with that in \oiii.  
\citet{seward05} give values for the X-ray flux based on the {\it Chandra} ACIS
data, viz.\ $F_{0.3-2.1\:keV} = 1.80 \times 10^{-10}\; \FLUX$, based on fits using a 
multi-component model with the absorption column density fixed at 
$N_H=6.16 \times 10^{21}\;{\rm cm^{-2}}$\@.  As we argue in \S4.1, a column of 
$3.3 \times 10^{21}\;{\rm cm^{-2}}$\ is more consistent all the data for G292.  It is not a 
simple matter to adjust the absorption, since changing $N_H$\ would change other 
parameters of the fit, but to first order we estimate that the lower absorption would
give an unabsorbed flux of $F_0 \approx  8 \times 10^{-10} \FLUX$, and an X-ray
luminosity $L_{0.3-2.1\: keV} \approx 3 \times 10^{36}\:d_6^{\:2}\; \LUM$, roughly 20 times the 
luminosity in \oiii \lam 5007\@.   Comparing G292 with the extraordinary SNR in NGC 4449, 
\citet{patnaude03} find for the latter $L_X = 2.4 \times 10^{36} \LUM$, just under half its 
luminosity in \oiii\@.
\footnote{The X-ray flux and luminosity given by \citet{patnaude03} are for energy range
0.5--2.1 keV,  slightly different from the one used by \citet{seward05} for G292.}
The \oiii/X-ray luminosity ratio for G292 is probably more
typical of O-rich SNRs, but it would be interesting to do a quantitative comparison for others.

\subsection{Filament Morphology---Rayleigh-Taylor Fingers}

The most striking feature of the outer filaments in G292 is that most of them display an elongated,
finger-like morphology directed generally outward from the center of the remnant.  
A common feature of two- and three-dimensional hydrodynamic models for core-collapse
supernovae is Rayleigh-Taylor instabilities, which form at interfaces between layers
of different composition as the supernova shock pushes outward through the outer core of 
the progenitor \citep[e.g.,][]{fryxell91, burrows95, nagataki98, kifonidis00, kifonidis03}.
These instabilities produce mixing of material from different layers,  
as observed in SN1987A \citep[e.g.,][]{arnett89, mccray93, wang02} and eventually 
may lead to SNR filaments of highly  
varied composition, as in Cas A \citep[e.g.,][]{chevalier78, reed95, hurford96, fesen01}.  
The geometry of most of the outer filaments 
in G292 is very suggestive of Rayleigh-Taylor instabilities: thin, irregular fingers oriented
radially, as illustrated by several examples shown in detail in Fig.\ 9, taken from locations 
indicated in Fig.\ 10.\footnote{\citet{ghavamian05} 
have suggested that the scalloped morphology of the 
bright spur filament may result from R-T instabilities that 
have resulted from the recent encounter of a reverse shock with an overdense shell of ejecta, 
and they have noted a similarity
to filaments in Cas A \citep{fesen01}.  
Our discussion here refers primarily to the outer filaments, whose finger-like geometry 
may be a remnant of R-T instabilities during the explosion.}
The arrows in each panel of the
figure are directed toward the geometric center of the outer radio shell, as defined by
\citet{gaensler03}:  $\alpha(2000) = 11^{\rm h}24^{\rm m}34\fs2,\ 
\delta (2000) = -59\degr 15\arcmin 54\arcsec$\@.   
\citep[The pulsar is  located about 45\arcsec\ 
to the SE relative to this center, but it could well be moving rapidly, as]
[have noted.]{hughes01}
These filaments could have originated as R-T fingers, formed when the shock pushed 
outward through the oxygen zone in the early stages of the supernova explosion, that have 
been ballistically expanding since.  The ejecta clumps would have only recently become visible,  
once they entered the reverse shock and became excited.   
\citet{blondin01} have shown that the
fingers should stay near the forward shock front throughout the time that the reverse shock is 
in the power-law portion of the ejecta density profile---which can extend up to several thousand
years.    

\notetoeditor{If possible we would like to have Figs. 9 and 10 on the same or facing pages
where they can easily be compared.}

\section{Summary}

We have obtained the first narrow-band CCD images to cover the entirety of the oxygen-rich
SNR G292.0+1.8 
and have also obtained a long-slit spectrum from a single particularly interesting position 
chosen for its strong contrast as seen in \oiii\ and \sii\ images.  Our findings can
be briefly summarized as follows:

1.  The \oiii\ filaments are widespread throughout much of the G292 shell, including 
a few very near the outer edge of the radio/X-ray shell.  All the filaments 
show no \ha\ emission and thus appear to be fragments of supernova ejecta,
the same conclusion reached by previous studies for filaments in the central region
of G292.  The filaments at the shell perimeter would seem to be bullets of ejecta that have
nearly caught up to the primary shock.  

2. We have discovered in our images a few filaments 
that emit in \sii\ in addition to oxygen lines, but that till show no \ha.  We have confirmed one of these
spectroscopically.  These \sii\ filaments represent the first evidence for 
products of oxygen burning in the fragments of pure supernova ejecta in G292.   
Clearly some oxygen burning must have occurred either hydrostatically in
the G292 progenitor or explosively during the supernova,
but the paucity of \sii\  filaments suggests either that O-burning was not extensive or that
most O-burning products have yet to be excited.

3. Most of the \oiii\ filaments, especially those outside the central region,
are radially oriented, pencil-like structures.  This is the structure that hydrodynamic
simulations for core-collapse supernovae show developing in the outer portion of the progenitor core
as a result of Rayleigh-Taylor instabilities.  We suggest that the present ejecta "bullets" 
seen optically may trace their origins to back to these features that developed during the explosion.

4. The extinction to G292 is probably equivalent to $E(B-V)=0.6$\ mag, 
or $A(V) = 1.9$\ mag, significantly less than has been used in previous discussions
of the optical emission from G292.  The lower extinction is more consistent both with
X-ray and radio measures for the hydrogen column density and with models for the
optical emission from a  shocked metal-rich gas.  

5. The integrated \oiii\ flux from G292 is $F_{5007} = 5.4 \times 10^{-12} \FLUX$.  
This value, with a distance of 6 kpc and our adopted value for the extinction,  implies 
$L_{5007} = 1.6 \times 10^{35} \LUM$, approximately 20 times smaller than the X-ray 
luminosity.  This is dramatically different from the extraordinary SNR in NGC 4449, 
which has an \oiii\ luminosity roughly twice that in X-rays.  
It would be interesting to obtain similar comparisons for other O-rich SNRs.   

The status of G292.0+1.8 as one of only a handful of oxygen-rich SNRs, and especially the
fact that it remains unique in displaying all the features expected to result from a core-collapse 
supernova, makes it an object of continuing interest for further study in all wavelength bands.

\acknowledgments

We gratefully acknowledge the outstanding support,  
typical of the mountain staff at CTIO, during the observations that yielded the 
new data reported here.    Several undergraduate students made important
contributions at various stages:  Becky Walldroff assisted in the acquisition 
of the spectroscopic data; Sarah Kate May carried out the initial spectroscopic data reduction;
and Claudine Reith assisted in acquiring and processing the image data.
Comments from Jack Hughes and from the anonymous referee 
have been valuable in preparing the final version of this paper.
This work has been made 
possible through the financial support from the NSF, through grant 
AST-0307613 to P.F.W., and from NASA, through grant NAG 5-8020 to P.F.W. and 
{\it Chandra} grants GO0-1120X and GO1-2058A to K.S.L.  Additional support for 
astrophysics research at Middlebury College has been provided by the W.M. 
Keck Foundation through the Keck Northeast Astronomy Consortium.

\bibliographystyle{apj}


\begin{thebibliography}{49}
\expandafter\ifx\csname natexlab\endcsname\relax\def\natexlab#1{#1}\fi

\bibitem[{{Arnett}(1996)}]{arnett96}
{Arnett}, D. 1996, {Supernovae and Nucleosynthesis} (Princeton, NJ: Princeton
  U. Press)

\bibitem[{{Arnett} {et~al.}(1989){Arnett}, {Bahcall}, {Kirshner}, \&
  {Woosley}}]{arnett89}
{Arnett}, W.~D., {Bahcall}, J.~N., {Kirshner}, R.~P., \& {Woosley}, S.~E. 1989,
  \araa, 27, 629

\bibitem[{{Blair} {et~al.}(1983){Blair}, {Kirshner}, \& {Winkler}}]{blair83}
{Blair}, W.~P., {Kirshner}, R.~P., \& {Winkler}, P.~F. 1983, \apj, 272, 84

\bibitem[{{Blair} {et~al.}(2000){Blair}, {Morse}, {Raymond}, {Kirshner},
  {Hughes}, {Dopita}, {Sutherland}, {Long}, \& {Winkler}}]{blair00}
{Blair}, W.~P., {Morse}, J.~A., {Raymond}, J.~C., {Kirshner}, R.~P., {Hughes},
  J.~P., {Dopita}, M.~A., {Sutherland}, R.~S., {Long}, K.~S., \& {Winkler},
  P.~F. 2000, \apj, 537, 667

\bibitem[{{Blair} {et~al.}(1989){Blair}, {Raymond}, {Danziger}, \&
  {Matteucci}}]{blair89}
{Blair}, W.~P., {Raymond}, J.~C., {Danziger}, J., \& {Matteucci}, F. 1989,
  \apj, 338, 812

\bibitem[{{Blondin} \& {Ellison}(2001)}]{blondin01}
{Blondin}, J.~M. \& {Ellison}, D.~C. 2001, \apj, 560, 244

\bibitem[{{Burrows} {et~al.}(1995){Burrows}, {Hayes}, \& {Fryxell}}]{burrows95}
{Burrows}, A., {Hayes}, J., \& {Fryxell}, B.~A. 1995, \apj, 450, 830

\bibitem[{{Camilo} {et~al.}(2002){Camilo}, {Manchester}, {Gaensler}, {Lorimer},
  \& {Sarkissian}}]{camilo02}
{Camilo}, F., {Manchester}, R.~N., {Gaensler}, B.~M., {Lorimer}, D.~R., \&
  {Sarkissian}, J. 2002, \apjl, 567, L71

\bibitem[{{Cardelli} {et~al.}(1989){Cardelli}, {Clayton}, \&
  {Mathis}}]{cardelli89}
{Cardelli}, J.~A., {Clayton}, G.~C., \& {Mathis}, J.~S. 1989, \apj, 345, 245

\bibitem[{{Chevalier}(2005)}]{chevalier05}
{Chevalier}, R.~A. 2005, \apj, 619, 839

\bibitem[{{Chevalier} \& {Kirshner}(1978)}]{chevalier78}
{Chevalier}, R.~A. \& {Kirshner}, R.~P. 1978, \apj, 219, 931

\bibitem[{{Dopita} \& {Tuohy}(1984)}]{dopita84}
{Dopita}, M.~A. \& {Tuohy}, I.~R. 1984, \apj, 282, 135

\bibitem[{{Fesen}(2001)}]{fesen01}
{Fesen}, R.~A. 2001, \apjs, 133, 161

\bibitem[{{Fryxell} {et~al.}(1991){Fryxell}, {Arnett}, \&
  {Mueller}}]{fryxell91}
{Fryxell}, B., {Arnett}, D., \& {Mueller}, E. 1991, \apj, 367, 619

\bibitem[{{Gaensler} \& {Wallace}(2003)}]{gaensler03}
{Gaensler}, B.~M. \& {Wallace}, B.~J. 2003, \apj, 594, 326

\bibitem[{{Gaustad} {et~al.}(2001){Gaustad}, {McCullough}, {Rosing}, \& {Van
  Buren}}]{gaustad01}
{Gaustad}, J.~E., {McCullough}, P.~R., {Rosing}, W., \& {Van Buren}, D. 2001,
  \pasp, 113, 1326

\bibitem[{{Ghavamian} {et~al.}(2005){Ghavamian}, {Hughes}, \&
  {Williams}}]{ghavamian05}
{Ghavamian}, P., {Hughes}, J.~P., \& {Williams}, T.~B. 2005, \apj, 635, 365

\bibitem[{{Gonzalez} \& {Safi-Harb}(2003)}]{gonzalez03}
{Gonzalez}, M. \& {Safi-Harb}, S. 2003, \apjl, 583, L91

\bibitem[{{Goss} {et~al.}(1979){Goss}, {Shaver}, {Zealey}, {Murdin}, \&
  {Clark}}]{goss79}
{Goss}, W.~M., {Shaver}, P.~A., {Zealey}, W.~J., {Murdin}, P., \& {Clark},
  D.~H. 1979, \mnras, 188, 357

\bibitem[{{Hamuy} {et~al.}(1992){Hamuy}, {Walker}, {Suntzeff}, {Gigoux},
  {Heathcote}, \& {Phillips}}]{hamuy92}
{Hamuy}, M., {Walker}, A.~R., {Suntzeff}, N.~B., {Gigoux}, P., {Heathcote},
  S.~R., \& {Phillips}, M.~M. 1992, \pasp, 104, 533

\bibitem[{{Hughes} \& {Singh}(1994)}]{hughes94}
{Hughes}, J.~P. \& {Singh}, K.~P. 1994, \apj, 422, 126

\bibitem[{{Hughes} {et~al.}(2001){Hughes}, {Slane}, {Burrows}, {Garmire},
  {Nousek}, {Olbert}, \& {Keohane}}]{hughes01}
{Hughes}, J.~P., {Slane}, P.~O., {Burrows}, D.~N., {Garmire}, G., {Nousek},
  J.~A., {Olbert}, C.~M., \& {Keohane}, J.~W. 2001, \apjl, 559, L153

\bibitem[{{Hughes} {et~al.}(2003){Hughes}, {Slane}, {Park}, {Roming}, \&
  {Burrows}}]{hughes03}
{Hughes}, J.~P., {Slane}, P.~O., {Park}, S., {Roming}, P.~W.~A., \& {Burrows},
  D.~N. 2003, \apjl, 591, L139

\bibitem[{{Hunter} {et~al.}(1999){Hunter}, {van Woerden}, \&
  {Gallagher}}]{hunter99}
{Hunter}, D.~A., {van Woerden}, H., \& {Gallagher}, J.~S. 1999, \aj, 118, 2184

\bibitem[{{Hurford} \& {Fesen}(1996)}]{hurford96}
{Hurford}, A.~P. \& {Fesen}, R.~A. 1996, \apj, 469, 246

\bibitem[{{Kifonidis} {et~al.}(2000){Kifonidis}, {Plewa}, {Janka}, \& {M{\"
  u}ller}}]{kifonidis00}
{Kifonidis}, K., {Plewa}, T., {Janka}, H.-T., \& {M{\" u}ller}, E. 2000, \apjl,
  531, L123

\bibitem[{{Kifonidis} {et~al.}(2003){Kifonidis}, {Plewa}, {Janka}, \&
  {M{\"u}ller}}]{kifonidis03}
{Kifonidis}, K., {Plewa}, T., {Janka}, H.-T., \& {M{\"u}ller}, E. 2003, \aap,
  408, 621

\bibitem[{{Kirshner} {et~al.}(1989){Kirshner}, {Morse}, {Winkler}, \&
  {Blair}}]{kirshner89}
{Kirshner}, R.~P., {Morse}, J.~A., {Winkler}, P.~F., \& {Blair}, W.~P. 1989,
  \apj, 342, 260

\bibitem[{{Long} {et~al.}(1989){Long}, {Blair}, \& {Krzeminski}}]{long89}
{Long}, K.~S., {Blair}, W.~P., \& {Krzeminski}, W. 1989, \apjl, 340, L25

\bibitem[{{McCray}(1993)}]{mccray93}
{McCray}, R. 1993, \araa, 31, 175

\bibitem[{{Mills} {et~al.}(1961){Mills}, {Slee}, \& {Hill}}]{mills61}
{Mills}, B.~Y., {Slee}, O.~B., \& {Hill}, E.~R. 1961, Australian Journal of
  Physics, 14, 497

\bibitem[{{Milne}(1969)}]{milne69}
{Milne}, D.~K. 1969, Australian Journal of Physics, 22, 613

\bibitem[{{Murdin} \& {Clark}(1979)}]{murdin79}
{Murdin}, P. \& {Clark}, D.~H. 1979, \mnras, 189, 501

\bibitem[{{Nagataki} {et~al.}(1998){Nagataki}, {Shimizu}, \&
  {Sato}}]{nagataki98}
{Nagataki}, S., {Shimizu}, T.~M., \& {Sato}, K. 1998, \apj, 495, 413

\bibitem[{{Park} {et~al.}(2004){Park}, {Hughes}, {Slane}, {Burrows}, {Roming},
  {Nousek}, \& {Garmire}}]{park04}
{Park}, S., {Hughes}, J.~P., {Slane}, P.~O., {Burrows}, D.~N., {Roming},
  P.~W.~A., {Nousek}, J.~A., \& {Garmire}, G.~P. 2004, \apjl, 602, L33

\bibitem[{{Park} {et~al.}(2002){Park}, {Roming}, {Hughes}, {Slane}, {Burrows},
  {Garmire}, \& {Nousek}}]{park02}
{Park}, S., {Roming}, P.~W.~A., {Hughes}, J.~P., {Slane}, P.~O., {Burrows},
  D.~N., {Garmire}, G.~P., \& {Nousek}, J.~A. 2002, \apjl, 564, L39

\bibitem[{{Patnaude} \& {Fesen}(2003)}]{patnaude03}
{Patnaude}, D.~J. \& {Fesen}, R.~A. 2003, \apj, 587, 221

\bibitem[{{Predehl} \& {Schmitt}(1995)}]{predehl95}
{Predehl}, P. \& {Schmitt}, J.~H.~M.~M. 1995, \aap, 293, 889

\bibitem[{{Rauscher} {et~al.}(2002){Rauscher}, {Heger}, {Hoffman}, \&
  {Woosley}}]{rauscher02}
{Rauscher}, T., {Heger}, A., {Hoffman}, R.~D., \& {Woosley}, S.~E. 2002, \apj,
  576, 323

\bibitem[{{Reed} {et~al.}(1995){Reed}, {Hester}, {Fabian}, \&
  {Winkler}}]{reed95}
{Reed}, J.~E., {Hester}, J.~J., {Fabian}, A.~C., \& {Winkler}, P.~F. 1995,
  \apj, 440, 706

\bibitem[{{Seward} {et~al.}(2005){Seward}, {Slane}, {Smith}, {Gaetz}, {Koo}, \&
  {Lee}}]{seward05}
{Seward}, F., {Slane}, P., {Smith}, R., {Gaetz}, T., {Koo}, B.-C., \& {Lee},
  J.-J. 2005, {Chandra Supernova Remnant Catalog},
  http://snrcat.cfa.harvard.edu/

\bibitem[{{Sutherland} \& {Dopita}(1995{\natexlab{a}})}]{sutherland95a}
{Sutherland}, R.~S. \& {Dopita}, M.~A. 1995{\natexlab{a}}, \apj, 439, 365

\bibitem[{{Sutherland} \& {Dopita}(1995{\natexlab{b}})}]{sutherland95b}
---. 1995{\natexlab{b}}, \apj, 439, 381

\bibitem[{{Tuohy} {et~al.}(1982){Tuohy}, {Burton}, \& {Clark}}]{tuohy82}
{Tuohy}, I.~R., {Burton}, W.~M., \& {Clark}, D.~H. 1982, \apjl, 260, L65

\bibitem[{{Wang} {et~al.}(2002){Wang}, {Wheeler}, {H{\"o}flich}, {Khokhlov},
  {Baade}, {Branch}, {Challis}, {Filippenko}, {Fransson}, {Garnavich},
  {Kirshner}, {Lundqvist}, {McCray}, {Panagia}, {Pun}, {Phillips}, {Sonneborn},
  \& {Suntzeff}}]{wang02}
{Wang}, L., {Wheeler}, J.~C., {H{\"o}flich}, P., {Khokhlov}, A., {Baade}, D.,
  {Branch}, D., {Challis}, P., {Filippenko}, A.~V., {Fransson}, C.,
  {Garnavich}, P., {Kirshner}, R.~P., {Lundqvist}, P., {McCray}, R., {Panagia},
  N., {Pun}, C.~S.~J., {Phillips}, M.~M., {Sonneborn}, G., \& {Suntzeff}, N.~B.
  2002, \apj, 579, 671

\bibitem[{{Winkler} \& {Kirshner}(1985)}]{winkler85}
{Winkler}, P.~F. \& {Kirshner}, R.~P. 1985, \apj, 299, 981

\bibitem[{{Winkler} {et~al.}(1991){Winkler}, {Roberts}, \&
  {Kirshner}}]{winkler91}
{Winkler}, P.~F., {Roberts}, P.~F., \& {Kirshner}, R.~P. 1991, in Supernovae.
  The Tenth Santa Cruz Workshop in Astronomy and Astrophysics, held July 9-21,
  1989, Lick Observatory. Editor, S.E. Woosley; Publisher, Springer-Verlag, New
  York, 1991. LC \# QB856 .S26 1989. ISBN \# 0387970711. P.652, 1991, 652--+

\bibitem[{{Woosley} {et~al.}(2002){Woosley}, {Heger}, \& {Weaver}}]{woosley02}
{Woosley}, S.~E., {Heger}, A., \& {Weaver}, T.~A. 2002, Reviews of Modern
  Physics, 74, 1015

\bibitem[{{Zacharias} {et~al.}(2000){Zacharias}, {Urban}, {Zacharias}, {Hall},
  {Wycoff}, {Rafferty}, {Germain}, {Holdenried}, {Pohlman}, {Gauss}, {Monet},
  \& {Winter}}]{ucac1_00}
{Zacharias}, N., {Urban}, S.~E., {Zacharias}, M.~I., {Hall}, D.~M., {Wycoff},
  G.~L., {Rafferty}, T.~J., {Germain}, M.~E., {Holdenried}, E.~R., {Pohlman},
  J.~W., {Gauss}, F.~S., {Monet}, D.~G., \& {Winter}, L. 2000, \aj, 120, 2131

\end{thebibliography}

\clearpage





\clearpage 

\begin{figure}
\figurenum{1a}
\plotone{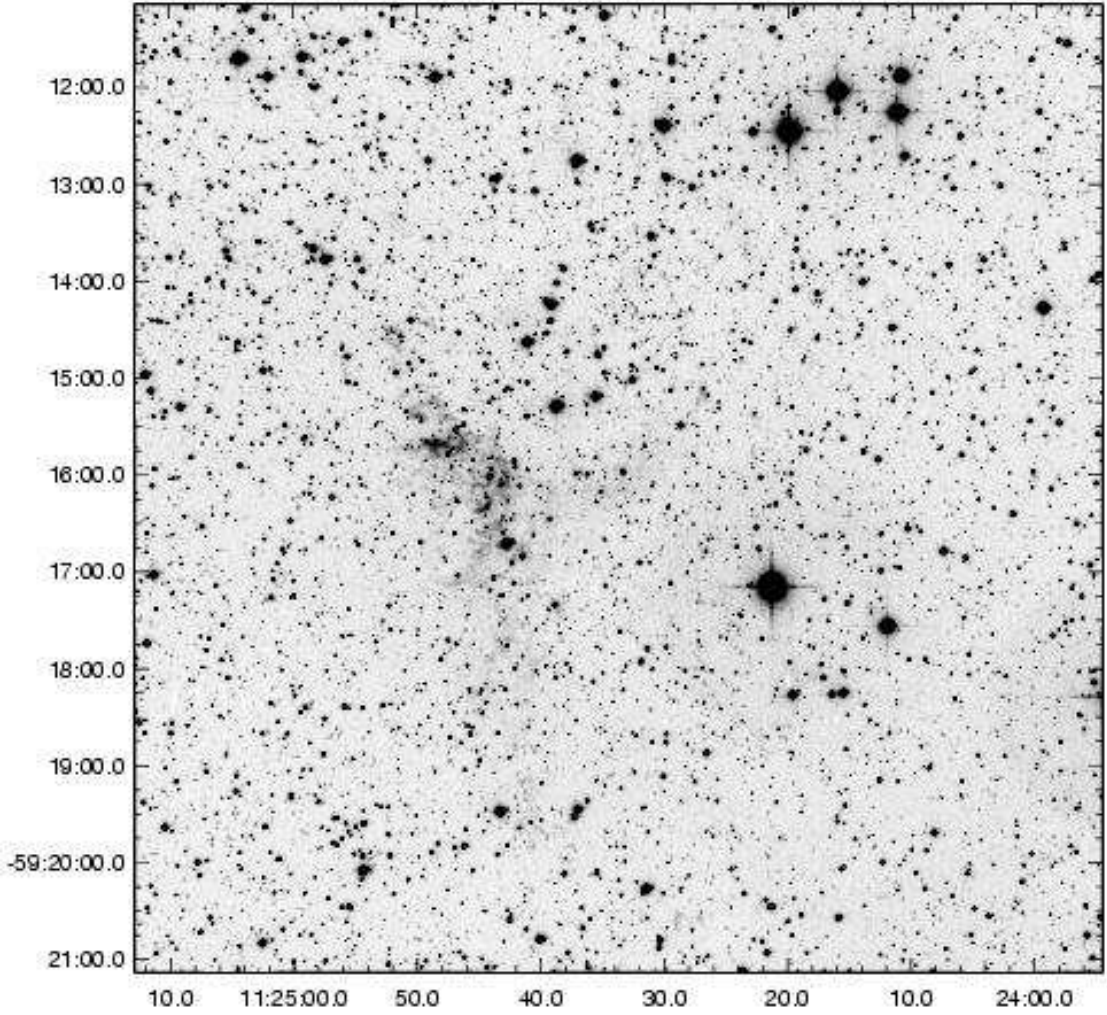}
\caption{SNR G292.0+1.8 in the light of \oiii.  The field is 10\arcmin\ square, with N up, and E  left, and is identical in Figures 1--4.}
\end{figure}

\clearpage

\begin{figure}
\figurenum{1b}
\plotone{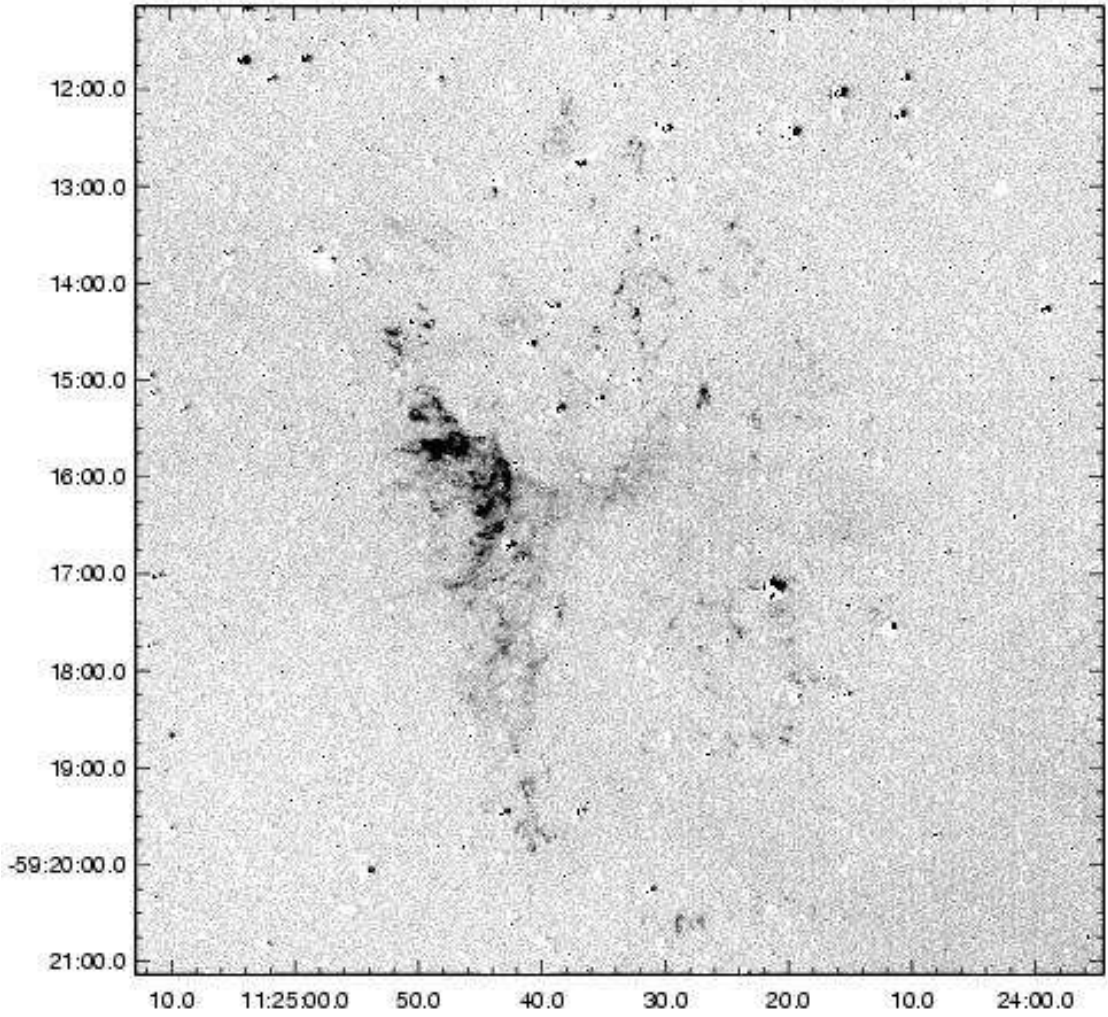}
\caption{SNR G292.0+1.8 in the light of \oiii\ (same image as Fig.\ 1a)  
with a matched continuum image subtracted to reveal the nebular features more clearly.}
\end{figure}

\clearpage



\begin{figure}
\figurenum{2a}
\plotone{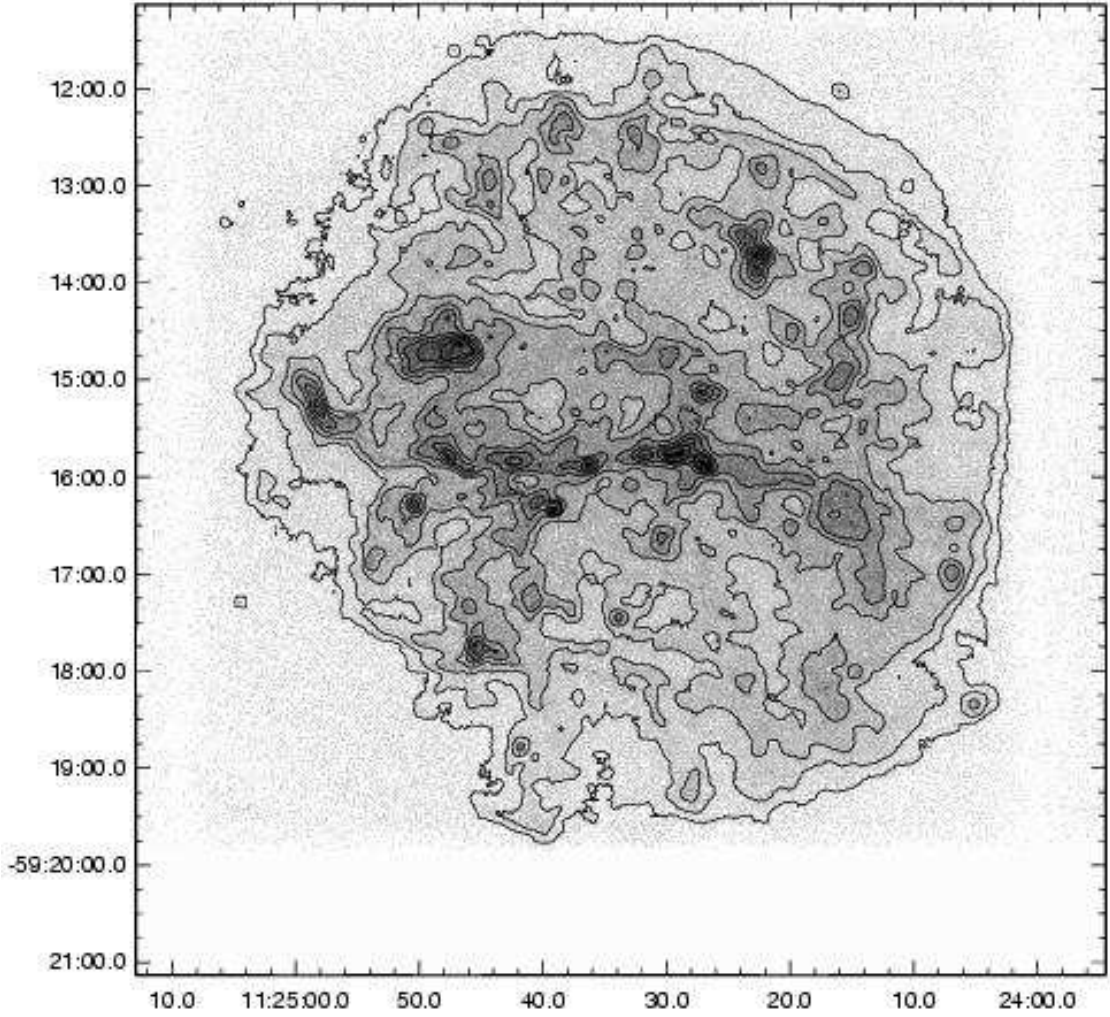}
\caption{{\it Chandra} ACIS image of the SNR G292.0+1.8 \citep{park02}, with contours added.  The field is identical to that in Fig.\ 1, but the southern edge of the ACIS chip is at $\delta(2000) \approx -59\degr 19\arcmin 45\arcsec$. }
\end{figure}
\clearpage

\begin{figure}
\figurenum{2b}
\plotone{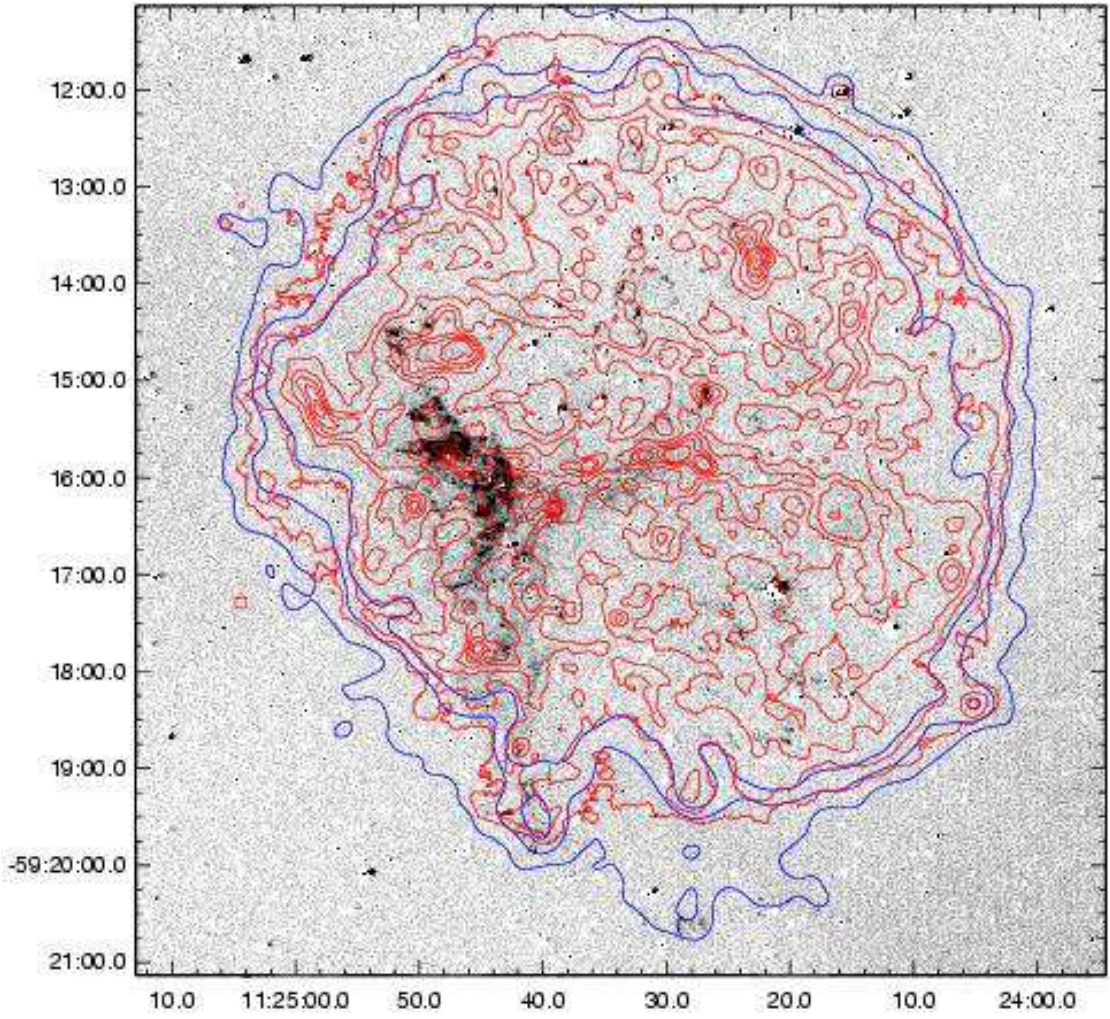}
\caption{The same continuum-subtracted \oiii\ image shown in Fig.\ 1b, with the ACIS contours shown in red.  The outermost contours of a ROSAT HRI image of G292 are also shown, in blue.}
\end{figure}
\clearpage

\begin{figure}
\figurenum{3}
\plotone{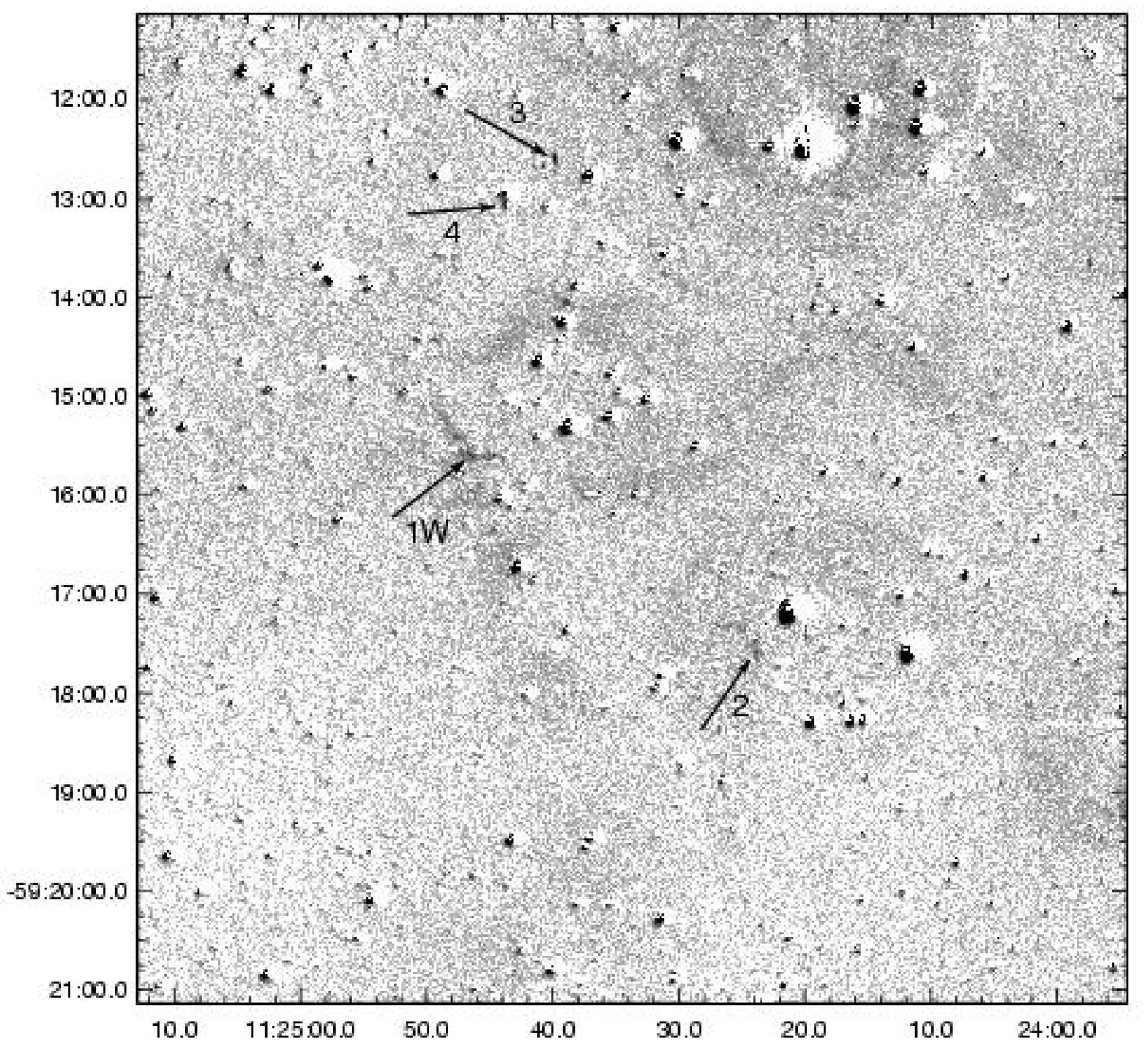}
\caption{SNR G292.0+1.8 in the light of \sii\ $\lambda\lambda $\  6716,6731, with a matched red continuum image subtracted.  The arrows indicate the locations of four groups of \sii\ filaments.   
Reflections from both the \sii\ and continuum filters has left significant residuals from bright stars.
The field is identical to that in Fig.\ 1.  
}
\end{figure}

\clearpage

\begin{figure}
\figurenum{4}
\plotone{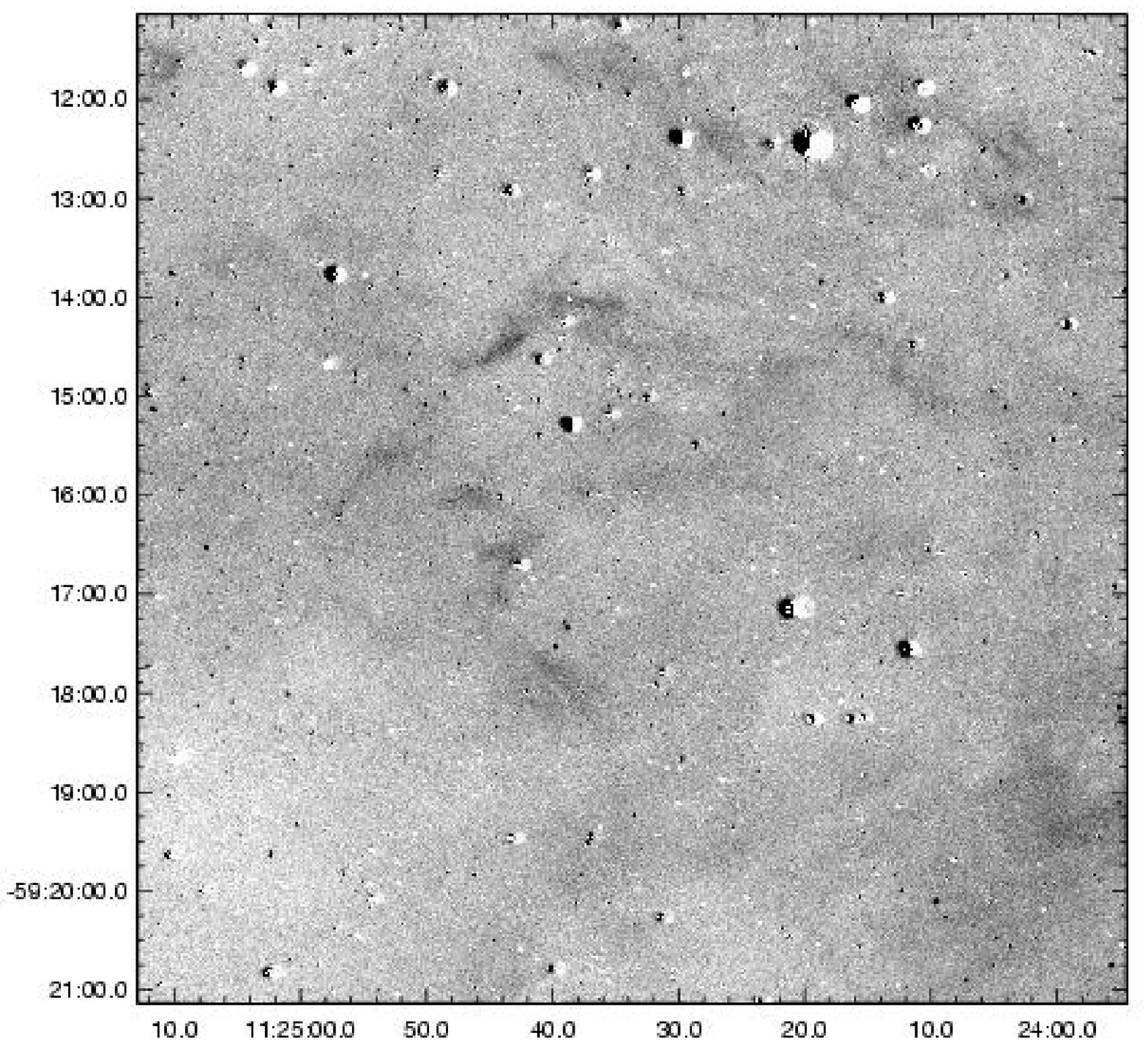}
\caption{SNR G292.0+1.8 in the light of \ha\  with a matched red continuum image subtracted.  Note the absence of any emission from the small filaments seen in \oiii\ and/or  \sii.   
The field is identical to that in Fig.\ 1.}
\end{figure}

\clearpage

\begin{figure}
\figurenum{5}
\plotone{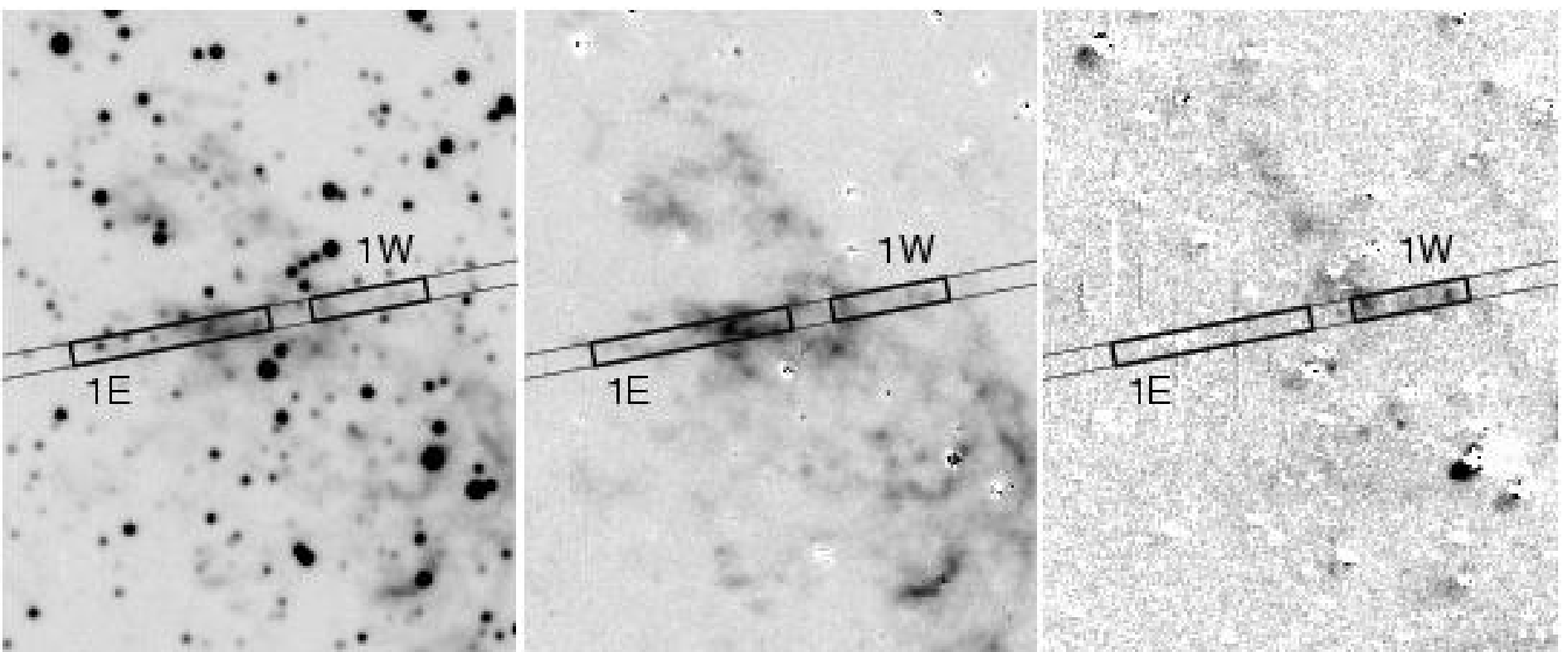}
\caption{The slit position for our long-slit spectrum (Fig.\ 6)  of G292.0+1.8 is shown on 
identical sections of three images:
({\it left}) \oiii, ({\it center}) continuum-subtracted \oiii, and ({\it right}) continuum-subtracted \sii.  
One-dimensional spectra (Fig.\ 7) were extracted from the two regions indicated by the 
heavier boxes:  1E includes the brightest \oiii\ filament in G292 (the northern portion of the ``Spur"): 
1W is a western extension of this filament and includes 
the brightest \sii\ filament known at the time our spectrum was obtained. 
The field shown is $80\arcsec \times 100\arcsec$.}
\end{figure}

\clearpage

\begin{figure}
\figurenum{6}
\plotone{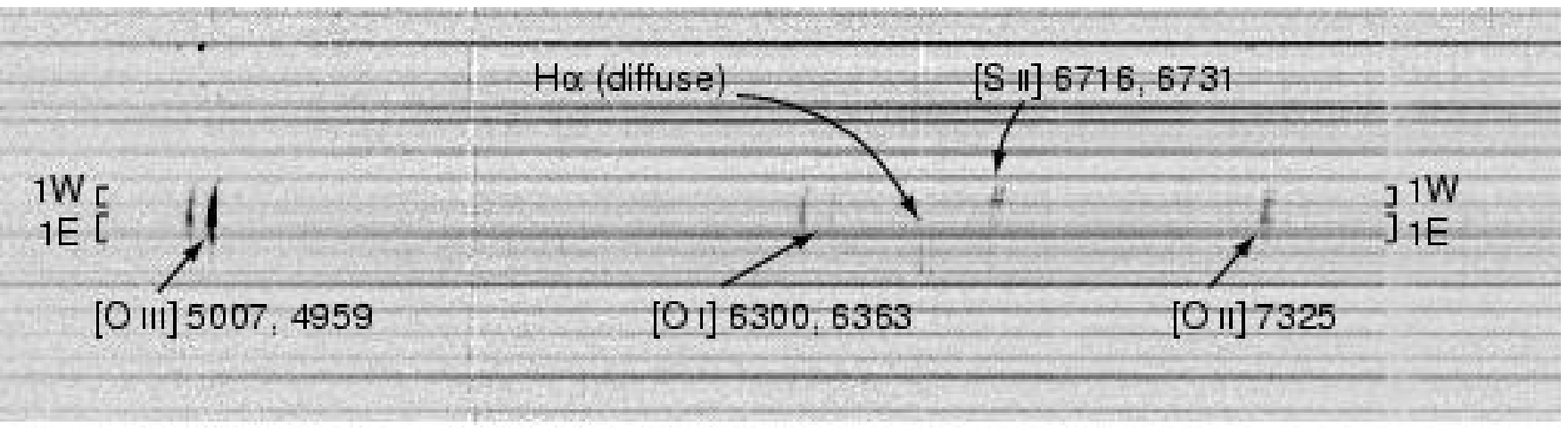}
\caption{Deep long-slit spectrum G292.0+1.8  with the slit oriented near 
east-west along the bright filament 1, as shown in Fig.\  5.  
This 2-D spectrum is from one of our 3 overlapping spectral ranges.  
The oxygen lines are all strongest in the main eastern portion (1E) of the filament, while 
\sii\  emission is strongest in the western extension (1W).  The filamentary emission shows 
a distinct curvature resulting from a radial velocity gradient of $\sim 500\kms$\ along the filament.  
The \ha\ emission shows no such velocity variation, and so has been largely removed in 
the sky subtraction Ð further indication that the filaments themselves are virtually devoid of H.
The two 1-D spectra shown in Fig.\ 7 were obtained by artificially removing the curvature,  
followed by extraction from regions 1E and 1W.
} 

\end{figure}

\clearpage

\begin{figure}
\figurenum{7}
\plotone{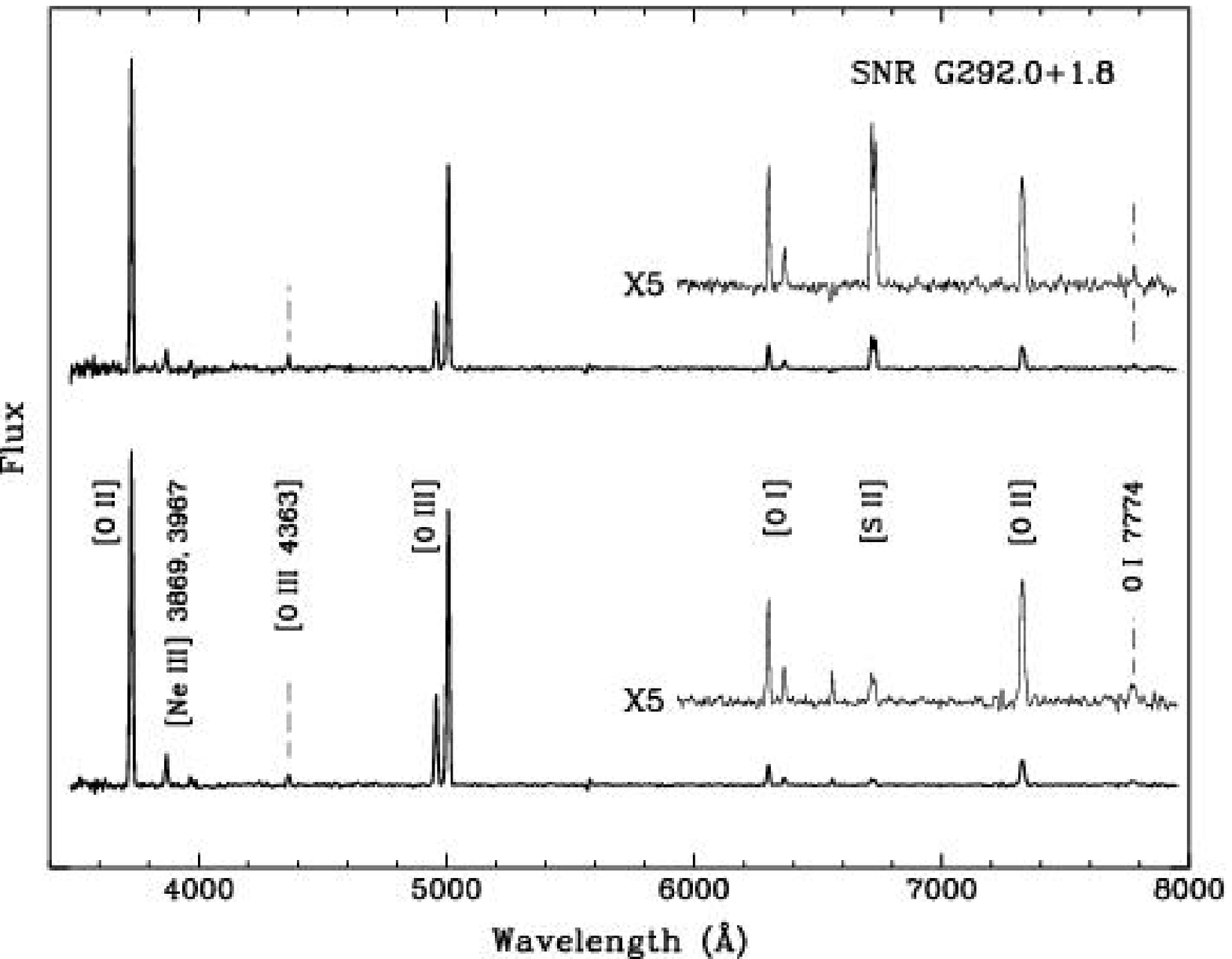}
\caption{These combined 1-D extracted spectra from two regions along the slit 
(filament 1E below, 1W above) show 
similar relative strengths from \oi, \oii, and \oiii, but the \sii\ lines  are 8 times stronger 
(relative to \oii) in the upper spectrum.  
Note also the faint line from permitted  \ion{O}{1}, which 
results from O$^+$\ winning out over H$^+$\  in recombination.  The long-wavelength ends of both spectra are repeated in the insets with the vertical 
scale magnified by a factor of 5.  
}
\end{figure}

\clearpage

\begin{figure}
\figurenum{8}
\plotone{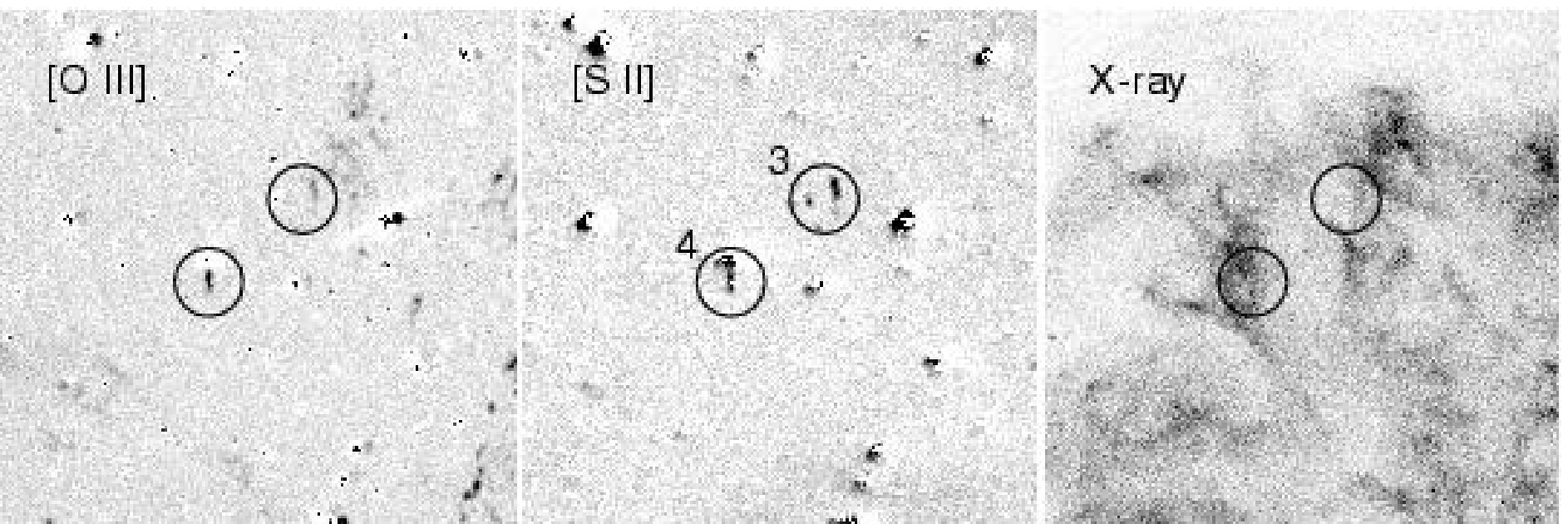}
\caption{This 2\farcm5 square section from the \oiii, \sii, and {\it Chandra} ACIS X-ray 
images of G292.0+1.8
shows the filament (No.\ 3) with the highest optical \sii/\oiii\ ratio.
In the same circle with filament 3 is 
another tiny S-rich knot,  8\arcsec\ to the east, which also has high \sii/\oiii\@.
Filament 4 coincides with an X-ray feature where \citet{park04} found Si and S to be 
more enhanced than anywhere else in G292.  
The circles have a diameter of 20\arcsec\ and the numbers on the \sii\ panel are the
same as in Fig.\ 3.
}
\end{figure}

\clearpage 

\begin{figure}
\figurenum{9}
\plotone{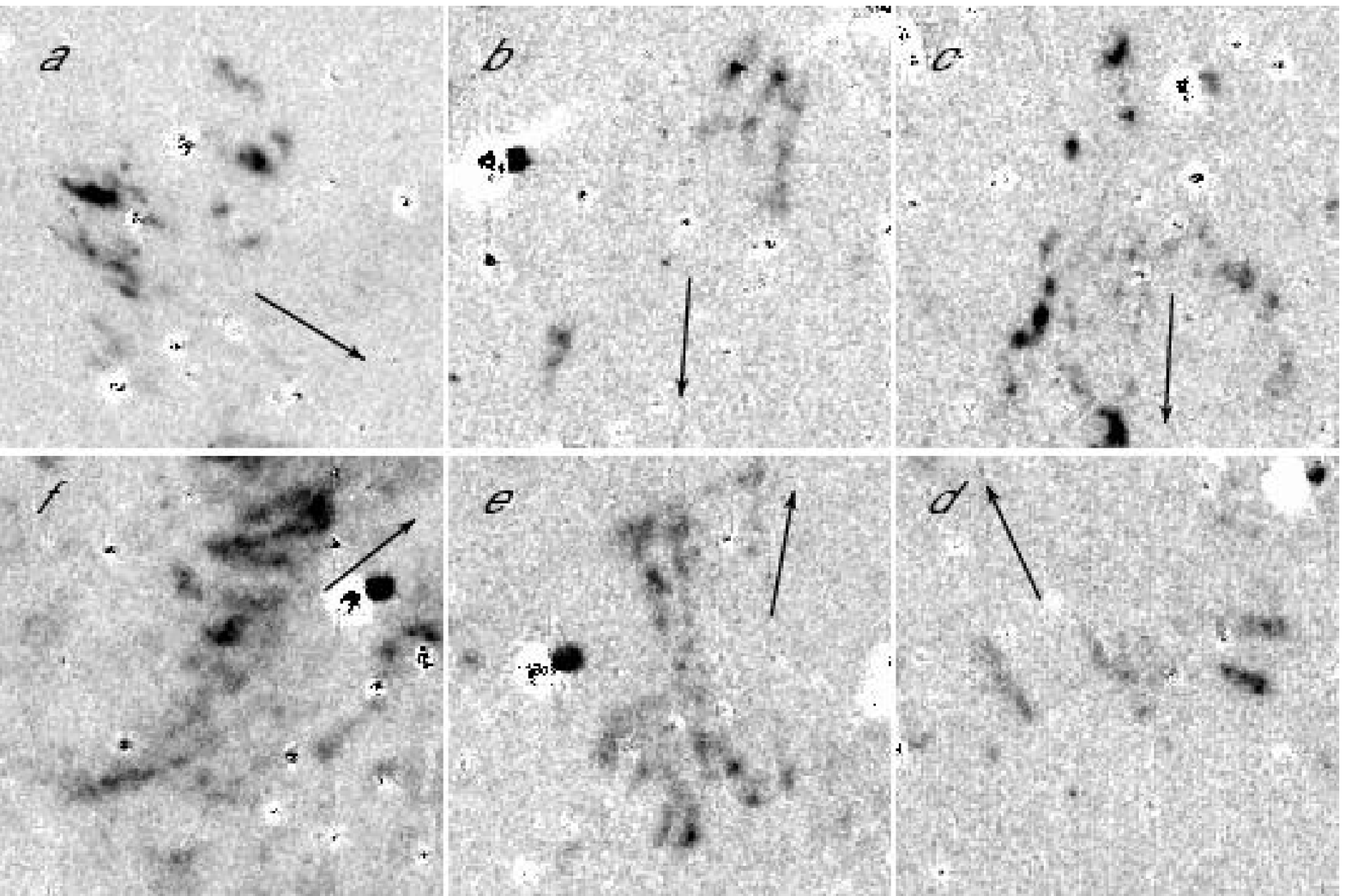}
\caption{These enlargements of  the continuum-subtracted \oiii\ image
(the same as that shown in Fig.\ 1b) 
show details of numerous ejecta filaments in G292.0+1.8, almost all
of which display a thin, pencil-like morphology that is suggests an origin as Rayleigh-Taylor 
fingers.   The arrows are all directed toward the geometric center of the radio shell, as 
given by \citet{gaensler03}; note that most of the fingers are oriented in a near-radial direction. 
Each of the panels is exactly 1\arcmin\ square, and are taken from locations {\it a-f} indicated in Fig.\ 10.
} 
\end{figure}

\clearpage

\begin{figure}
\figurenum{10}
\plotone{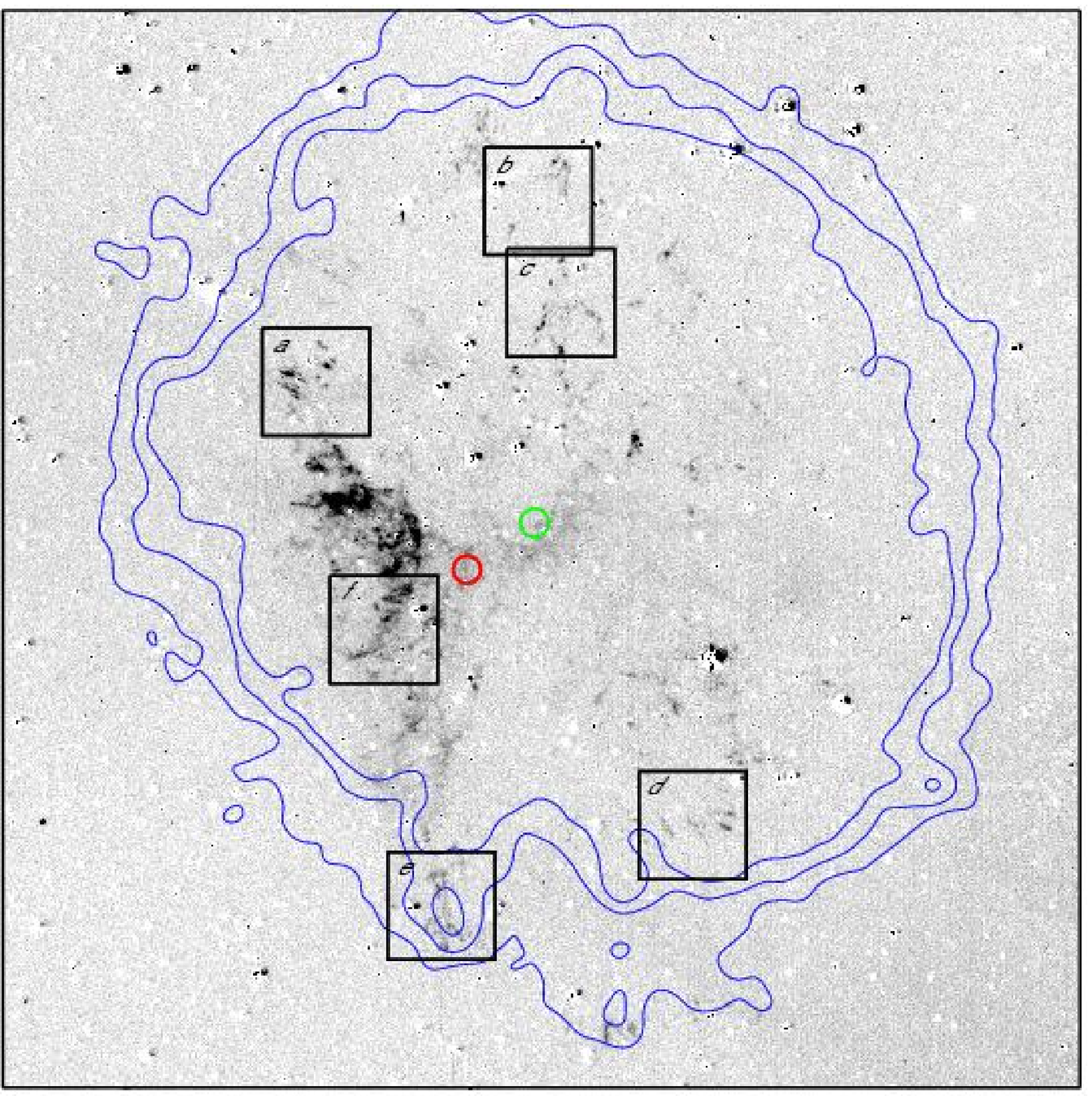}
\caption{The location from which each of the panels {\it a-f} in Fig.\ 9 was taken is
indicated by the corresponding field on this continuum-subtracted \oiii\ image of G292.0+1.8.
The red circle indicates the position of the pulsar PSR J1124--5916,
and the green circle indicates
the geometric center of the radio shell as given by \citet{gaensler03}.  
As an additional reference, the outer X-ray contours 
(from the {\it ROSAT} HRI, the same as shown in Fig.\ 2b) are shown in blue.
}
\end{figure}

\clearpage

\begin{deluxetable}{lccrcrc}
\tablecaption{Imaging Observations of G292.0+1.8. }
\tablewidth{0pt}
\tablehead{
\colhead{} & \multicolumn{2}{c}{Filter} &   \multicolumn{3}{c}{Velocity Range\tablenotemark{b} }&
\colhead{Exposure}   \\
\colhead{Designation} & 
\colhead{$\lambda_0$\ (\AA)} &\colhead{$\Delta\lambda$\tablenotemark{a}\ (\AA)}  & 
\multicolumn{3}{c}{$\kms$} &\colhead{ (s)} 
}
\tablenotetext{a}{FWHM.}  
\tablenotetext{b}{For \sii, the velocity range is given separately for  the 6716 \AA\  and 6731 \AA\ lines.}
\startdata
\oiii\ \lam5007 &5008  & \phn58 & --1700& - &+1800 & $5  \times 1000 $ \\
Green Continuum &5133  & 100 & &&& $5 \times \phn500 $ \\
\ha &6565  &  \phn24 & --\phn450 & - &+\phn650&$3  \times 1000 $ \\
\sii\ \lamlam 6716, 6731 &6728  &  \phn48 & --\phn550& - &+1600 & $4  \times 1000 $ \\
& & & --1200 & - &+\phn950 \\
Red Continuum &6848  &  \phn94 &&& & $5  \times \phn400 $ \\
\enddata


\end{deluxetable}

\clearpage

\begin{deluxetable}{lccccc}
\tablecaption{Spectroscopic Observations of G292.0+1.8 }
\tablewidth{0pt}
\tablehead{
\colhead{} & {} &  {} & {Wavelength Coverage} & {} & {Exposure}   \\
\colhead{Setup} & {Grating} & {Tilt} & (\AA) & Filter & (s) 
}
\startdata
Blue &09  & 13\fdg27 & 3,500--6,900 & --- & $4  \times 1800 $ \\
Red &32  & 14\fdg10 & 4,560--7,960 & GG420 & $5  \times 1200 $ \\
IR & 32  & 15\fdg66 & 6,600--10,000 & RG610 & $4  \times 1800 $ \\

\enddata


\end{deluxetable}

\clearpage 

\begin{deluxetable}{ccrrcrrcrr}
\tabletypesize{\scriptsize}
\tablecaption{Observed Emission-Line Fluxes in G292.0+1.8 }
\tablewidth{0pt}
\tablehead{
\colhead{} & {} &  \multicolumn{2}{c}{Flux\tablenotemark{a,b}\ (\taboiii\ \lam 5007 = 100)}  
& \phm{nn} &  \multicolumn{2}{c}{Intensity (\taboiii \lam 5007 = 100)} 
& \phm{nn} &  \multicolumn{2}{c}{Intensity (\taboiii\ \lam 5007 = 100)} \\
\colhead{Wavelength}&{}&{}&{}&{}
&\multicolumn{2}{c}{Dereddened,\tablenotemark{c} $E(B-V)=0.6$}&{}
&\multicolumn{2}{c}{Dereddened,\tablenotemark{c} $E(B-V)=0.9$}\\
\cline{3-4} \cline{6-7} \cline{9-10}
\colhead{(\AA)} & {Ion} & {Fil. 1 East} & {Fil. 1 West} & &  {Fil. 1 East} & {Fil. 1 West}
& &  {Fil. 1 East} & {Fil. 1 West}
}
\tablenotetext{a}{For Fil. 1 East, $F_{5007} = 10.8 \times 10^{-14}  \FLUX$, summed 
along 31\arcsec\  of the 3\farcs 5 slit; for Fil. 1 West, $F_{5007} = 3.5 \times 10^{-14}  \FLUX$, 
summed along 19\arcsec\ of the same slit. }
\tablenotetext{b}{Values in parentheses have high uncertainty.}
\tablenotetext{c}{The extinction function of
\citet{cardelli89} with $A(V)/E(B-V) = 3.1$\ was used for the dereddening.}  
\startdata

3727 & \taboii  & 117\phd\phn & 147\phd\phn & & 240\phd\phn & 302\phd\phn & 
&343\phd\phn & 432\phd\phn \\
3869 & \tabneiii  & 10\phd\phn & 9\phd\phn  & & 20\phd\phn & 18\phd\phn & 
& 28\phd\phn & 25\phd\phn \\
3967 & \tabneiii  & 2.5 & 3.5 & & 4.5 & 6.4 & & 6.2 &8.5 \\
4363 & \taboiii  & 3.6 & 5.3 & & 5.3 & 7.8 & & 6.4 & 9.4 \\
4959 & \taboiii  & 33\phd\phn & 32\phd\phn & & 34\phd\phn & 33\phd\phn & 
& 34\phd\phn & 33\phd\phn  \\
5007 & \taboiii  & 100.0 & 100.0  & & 100.0 & 100.0 & & 100.0 & 100.0\\

6300 & \taboi  & 6.9 & 12 & & 4.5 & 7.6 & & 3.6 & 6.0 \\
6363 & \taboi  & 2.6 & 3.9 & & 1.7 & 2.5 & & 1.3 & 2.0 \\
6563 & \ha  & $< 2\phd\phn$ & $< 1\phd\phn$  & & $< 1\phd\phn$ & $< 1\phd\phn$ & 
& $< 1\phd\phn$ & $< 1\phd\phn$\\

6716 & \tabsii  & 2.0 & 17\phd\phn & & 1.2 & 9.4 & & 0.9 & 7.1 \\
6731 & \tabsii  & 1.3 & 14\phd\phn & & 0.8 & 8.1 & & 0.6  & 6.2 \\
7325 & \taboii  & 14\phd\phn & 17\phd\phn & & 6.7 & 8.4  & & 4.7 & 5.9\\
7774 & O\thinspace I  & (2)\phn & (2)\phn & & (0.7) & (0.8) & & (0.5) & (0.5) \\

\enddata


\end{deluxetable}

\clearpage

\end{document}